\documentclass[twocolumn]{aastex61}

\usepackage{graphicx,amsmath,amsfonts,amssymb,subfigure,natbib,hyperref,rotating}
\newcommand{\fsps}{\texttt{FSPS}}

\newcommand{\pythonfsps}{\texttt{python-fsps}}

\newcommand{\mname}{\texttt{Prospector-$\alpha$}}

\newcommand{\prospector}{\texttt{Prospector}}

\newcommand{\dustone}{$\hat{\tau}_{\lambda,1}$}
\newcommand{\dusttwo}{$\hat{\tau}_{\lambda,2}$}
\newcommand{\didx}{n}

\newcommand{\fagn}{f$_{\mathrm{AGN}}$}

\newcommand{\tauagn}{$\tau_{\mathrm{AGN}}$}

\newcommand{\dynesty}{\texttt{dynesty}}

\newcommand{\tuniv}{$t_{\mathrm{univ}}$}
\newcommand{\sfruvir}{SFR$_{\mathrm{UV+IR}}$}

\newcommand{\lir}{{L$_{\mathrm{IR}}$}}
\newcommand{\luv}{{L$_{\mathrm{UV}}$}}

\begin{document}

\title{An Older, More Quiescent Universe from Panchromatic SED Fitting of the 3D-HST Survey}

\correspondingauthor{Joel Leja}
\email{joel.leja@cfa.harvard.edu}

\author[0000-0001-6755-1315]{Joel Leja}
\affil{Harvard-Smithsonian Center for Astrophysics, 60 Garden St. Cambridge, MA 02138, USA}
\affil{NSF Astronomy and Astrophysics Postdoctoral Fellow}

\author[0000-0002-9280-7594]{Benjamin D. Johnson}
\affil{Harvard-Smithsonian Center for Astrophysics, 60 Garden St. Cambridge, MA 02138, USA}

\author[0000-0002-1590-8551]{Charlie Conroy}
\affil{Harvard-Smithsonian Center for Astrophysics, 60 Garden St. Cambridge, MA 02138, USA}

\author[0000-0002-8282-9888]{Pieter van Dokkum}
\affil{Department of Astronomy, Yale University, New Haven, CT 06511, USA}

\author[0000-0003-2573-9832]{Joshua S. Speagle}
\affil{Harvard-Smithsonian Center for Astrophysics, 60 Garden St. Cambridge, MA 02138, USA}

\author[0000-0003-2680-005X]{Gabriel Brammer}
\affil{University of Copenhagen, N\o{}rregade 10, 1165 K\o{}benhavn, Denmark}

\author[0000-0003-1665-2073]{Ivelina Momcheva}
\affil{Space Telescope Science Institute, 3700 San Martin Drive, Baltimore, MD 21211 USA}

\author[0000-0001-7393-3336]{Rosalind Skelton}
\affil{South African Astronomical Observatory, PO Box 9, Observatory, Cape Town 7935, South Africa}

\author[0000-0001-7160-3632]{Katherine E. Whitaker}
\affil{Department of Physics, University of Connecticut, 2152 Hillside Road, Unit 3046, Storrs, CT 06269, USA}

\author[0000-0002-8871-3026]{Marijn Franx}
\affil{Leiden Observatory, Leiden University, NL-2300 RA Leiden, Netherlands}

\author[0000-0002-7524-374X]{Erica J. Nelson}
\affil{Harvard-Smithsonian Center for Astrophysics, 60 Garden St. Cambridge, MA 02138, USA}

\submitjournal{ApJ}
\begin{abstract}
Galaxy observations are influenced by many physical parameters: stellar masses, star formation rates (SFRs), star formation histories (SFHs), metallicities, dust, black hole activity, and more. As a result, inferring accurate physical parameters requires high-dimensional models which capture or marginalize over this complexity. Here we re-assess inferences of galaxy stellar masses and SFRs using the 14-parameter physical model \mname{} built in the \prospector{} Bayesian inference framework. We fit the photometry of 58,461 galaxies from the 3D-HST catalogs at $0.5 < z < 2.5$. The resulting stellar masses are $\sim0.1-0.3$ dex larger than the fiducial masses while remaining consistent with dynamical constraints. This change is primarily due to the systematically older SFHs inferred with \prospector{}. The SFRs are $\sim0.1-1+$ dex lower than UV+IR SFRs, with the largest offsets caused by emission from ``old'' ($t>100$ Myr) stars. These new inferences lower the observed cosmic star formation rate density by $\sim0.2$ dex and increase the observed stellar mass growth by $\sim 0.1$ dex, finally bringing these two quantities into agreement and implying an older, more quiescent Universe than found by previous studies at these redshifts. We corroborate these results by showing that the \mname{} SFHs are both more physically realistic and are much better predictors of the evolution of the stellar mass function. Finally, we highlight examples of observational data which can break degeneracies in the current model; these observations can be incorporated into priors in future models to produce new \& more accurate physical parameters.
\end{abstract}

\section{Introduction}
The modern approach to galaxy spectral energy distributions (SEDs) with stellar population synthesis (SPS) models was pioneered by \citet{sawicki98}. These authors fit the rest-frame UV-optical broadband photometry of Lyman-break galaxies with an exponentially declining $\tau$-model SFH, allowing variation in the start time, the duration of star formation ($\tau$), the stellar metallicity, and a reddening factor. This basic formula of a 4-5 parameter model covering a simple functional SFH, a dust attenuation vector, and perhaps stellar metallicity has remained a robust feature in the literature over the past two decades \citep{brinchmann00,papovich01,shapley01,ilbert06,salim07,kriek09,maraston10,acquaviva11,skelton14,salmon15}.
 
These fits have been extraordinarily successful as they provide a physical map from galaxy photometry to physical properties. The most widely used parameters from such fits are star formation rates and stellar masses (e.g., \citealt{shapley01,hopkins06,madau14,genel14,speagle14,behroozi19}). Stellar masses are considered particularly robust due to fortuitous degeneracies between dust, age, and metallicity, which means that there is a fairly tight relation between M/L ratio and color \citep{bell01}.

However, there are known uncertainties and systematic errors in this approach. There has remained a persistent and systematic factor of two uncertainty in stellar masses derived from SED fitting codes \citep{papovich01,marchesini09,wuyts09,behroozi10,pforr12,conroy13a,mitchell13,mobasher15,santini15,leja15,tomczak16,leja19,carnall18}, while star formation rates (SFRs) obtained via either monochromatic indicators or SED modeling are subject to similar $0.3-0.5$ dex systematics \citep{wuyts11a,speagle14,carnall18,leja18}. These systematics are caused by a combination of: (1) fundamental uncertainties in the input physics such as dust models, stellar evolution, initial mass function (IMF), and stellar spectral libraries, and (2) observations which are at best weakly informative about the complexities of extragalactic stellar populations, resulting in strong model degeneracies. Examples of specific issues include differences in the underlying physics of SPS models ($\sim$0.1-0.2 dex), degeneracies from fundamental limitations such as the ``outshining'' of old stellar populations by young stars, the relative similarity of old stellar populations, and the age-dust-metallicity degeneracy (for a more complete list, see the review by \citealt{conroy13a} and discussion therein). Due to the many confounding factors, solving any one of these problems in isolation is challenging and requires very carefully designed experiments (e.g. measuring contribution of TP-AGB stars to the near-IR fluxes, \citealt{kriek10}). As a result the conventional wisdom has been that there is a nigh-unbreakable factor-of-two error in SED fitting outputs. This has created little incentive to improve on the basic SED fitting approach presented in \citet{sawicki98}, which is likely related to the persistence of this 4-5 parameter framework in the literature.

Fortunately, many big-picture questions in galaxy evolution are on order-of-magnitude scales and relatively insensitive to uncertainties at the factor of two level. For example, the cosmic star formation rate density is now known to peak at $z\sim2$ \citep{madau14}, the amount of stellar mass in the Universe has increased by a factor of $\sim$4 since $z\sim2$ \citep{madau14}, and galaxies likely reionized the Universe around $z\sim7$ \citep{schmidt14,mason18}.
 
However, our understanding of many other key aspects of galaxy formation is sensitive to factor of two systematics in stellar mass, star formation rates, and other SED fitting parameters. Massive galaxies are thought to approximately double their stellar mass from $z=2$ to the present \citep{vd10,patel13a} while Milky Way-mass galaxies grow their mass by a factor of $\sim10$ \citep{vd13,patel13b,papovich15}. Both star-forming and quiescent galaxies approximately double their size at a fixed stellar mass from $z=2.75$ \citep{vdw14}. The stellar mass--metallicity relationship most likely evolves at fixed mass by a factor of $\sim2$ from $z\sim2$ to the present in observations \citep{erb06a} and simulations \citep{torrey19}. A fundamental factor of two uncertainty in stellar mass means that even well-measured dynamical masses cannot be used to constrain the dark matter fraction in the inner regions of a galaxy \citep{cappellari12,vandesande15,wuyts16}. The slope of the star-forming sequence is quite sensitive to factor-of-two changes (e.g., \citealt{shivaei17}), meaning that relatively small changes in this slope can cause large changes in inferred galaxy formation histories \citep{leitner12,leja15} or that gas depletion times are no longer constant \citep{genzel15}. Systematic factor-of-two changes in SED-derived parameters can invalidate or inalterably change any or all of these conclusions. This presents a strong motivation to break the ``factor of two barrier'', the same motivation which has inspired our new approach to galaxy SED fitting.

Fortunately, many of the model improvements needed for this work have seen significant improvement over the past several decades. MAGPHYS was the first code to use energy balance to tie together UV-NIR and MIR-FIR photometry into a single physical model \citep{dacunha08}. More complex and more flexible star formation history parameterizations have been explored, starting with SFH libraries with random bursts superimposed \citep{kauffmann03a,gallazzi05,dacunha08}, to fits using multiple parametric SFHs \citep{iyer17,lee18}, to nonparametric piecewise-constant SFHs \citep{fernandes05,ocvirk06,tojeiro07,leja17}, to libraries of SFHs from simulations \citep{finlator07,pacifici12}. Spatially complex dust attenuation models have been developed which include extra attenuation towards younger star-forming regions \citep{charlot00} and flexible attenuation curves \citep{noll09,salmon16,leja17,salim18}. Emission from central active galactic nuclei (AGN) is now built into many SED fitting models \citep{berta13,ciesla15b,rivera16,leja18}. Including the effect of nebular emission using photoionization models such as CLOUDY \citep{ferland99,ferland13} and MAPPINGS III \citep{groves04} has become standard practice. Large uncertainties in the IR contribution of TP-AGB stars have largely been resolved \citep{maraston06,kriek10}, though other fundamental uncertainties in stellar population synthesis techniques remain (e.g. the effect of binaries and rotation on the ionizing flux production rates of massive stars, \citealt{choi17}).

These new model components necessitate more robust statistical frameworks to properly constrain them. Bayesian forward-modeling techniques pioneered by \citet{kauffmann03a}; \citet{burgarella05}; and \citet{salim07} help to constrain the complex, correlated parameter uncertainties typically present in galaxy models. Classic grid-based models grow exponentially in size with model dimensionality, but gridless `on-the-fly' models combined with Markov chain Monte Carlo algorithms can efficiently explore high-dimensional ($N\gtrsim7$) spaces \citep{chevallard16,leja17,carnall18}. The computational time necessary for on-the-fly model exploration is both less expensive and more readily available than ever before.

By combining many of these advances into a single consistent framework, it may be possible to finally break the factor of two accuracy barrier in galaxy SED modeling. Here we take the first step towards this goal with the \mname{} physical model built within the \prospector{} inference framework. \mname{} has been cross-calibrated by fitting broadband photometry and using the posteriors to predict independent spectroscopic and spatially resolved data as an external check \citep{leja17,leja18}. These checks ensure that SED fitting results are consistent with the overall picture of galaxy formation; given the lack of ``ground truth'' in SED modeling, such an approach is necessary to ensure accurate results. This necessitates an iterative cycle of refining the model, fitting new data, performing new predictive checks, and further refining the model. These new data could include large catalogs of photometry at longer wavelengths from e.g. ALMA or $Herschel$, or intermediate-redshift information-rich spectroscopic surveys such as as MOSDEF or KMOS-3D \citep{kriek15,wisnioski15}. This approach sets us on a long path, but it is the best path available to move the field forward.

This model is fit to the 3D-HST photometric catalogs. These are ideal data to investigate the population-wide 0.3 dex systematic errors in SED fitting: they provide rest-frame UV-IR photometry for $\sim$180,000 galaxies across $0.5 < z < 2.5$ and are complete in stellar mass down to $\sim10^9$ M$_{\odot}$ at $z=2$ \citep{tal14}.

Section \ref{sec:catalogs} describes the 3D-HST catalogs and how they are fit. Section \ref{sec:sedmodel} describes the SED model that is fit to these photometry. Section \ref{sec:deltased} details how the \mname{} masses and SFRs differ from previous estimates. Section \ref{sec:cross_validation} performs model cross-validation tests to explore the accuracy of the inferred parameters and also shows the change in the cosmic star formation rate density (SFRD) as a result of the new measurements. The results and next steps are discussed in Section \ref{sec:discussion} and the conclusion is presented in Section \ref{sec:conclusion}. This work is done with a \citet{chabrier03} IMF and a WMAP9 cosmology \citep{hinshaw13}. Unless otherwise noted, all parameters are reported as the median of the posterior probability distribution function (PDF).

\section{Sample and Data}
\label{sec:catalogs}
Galaxies are selected from the 3D-HST photometric catalogs \citep{skelton14}. The 3D-HST catalogs consist of state-of-the-art PSF-matched UV-IR photometry for hundreds of thousands of distant galaxies, covering $\sim$900 arcmin$^2$ in five well-studied extragalactic fields. Galaxies are identified in deep near-infrared Hubble Space Telescope (HST) imaging from the CANDELS survey \citep{grogin11,koekemoer11} and are covered by between 17 (the UDS field) to 44 (the COSMOS field) photometric bands spanning a range of 0.3-8$\mu$m in the observed frame. The photometry is supplemented by $Spitzer$/MIPS 24$\mu$m fluxes from \citet{whitaker14}. The MIPS 24$\mu$m coverage is critical because the rest-frame MIR wavelengths are dominated by warm dust emission, a key empirical proxy for obscured star formation \citep{kennicutt98}. Obscured star formation is the dominant form of star formation for massive galaxies in this redshift range \citep{whitaker17}. 

The 3D-HST catalogs contain additional stellar populations parameters, including stellar masses from FAST \citep{kriek09} and \sfruvir{} \citep{whitaker14}. In this work these parameters are referred to as the 3D-HST catalog masses and SFRs. The photometry is complete in stellar mass to at least $M_*$ = 10$^{9.3}$ M$_{\odot}$ between $0.5 < z < 2.5$ \citep{tal14}. Redshifts are taken from, in order of reliability: (1) ground-based spectroscopic redshifts, (2) near-infrared grism redshifts from the 3D-HST survey \citep{momcheva16}, and (3) photometric redshifts from EAZY \citep{brammer08,skelton14}.

\subsection{Sample selection}
\begin{figure*}[t!h!]
\begin{center}
\includegraphics[width=0.95\linewidth]{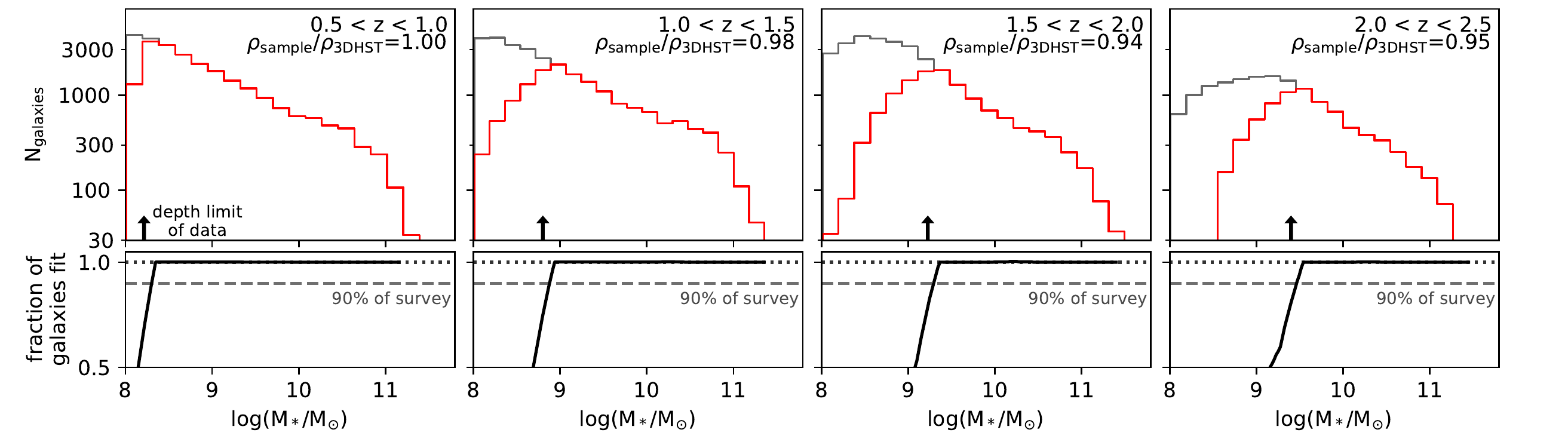}
\includegraphics[width=0.95\linewidth]{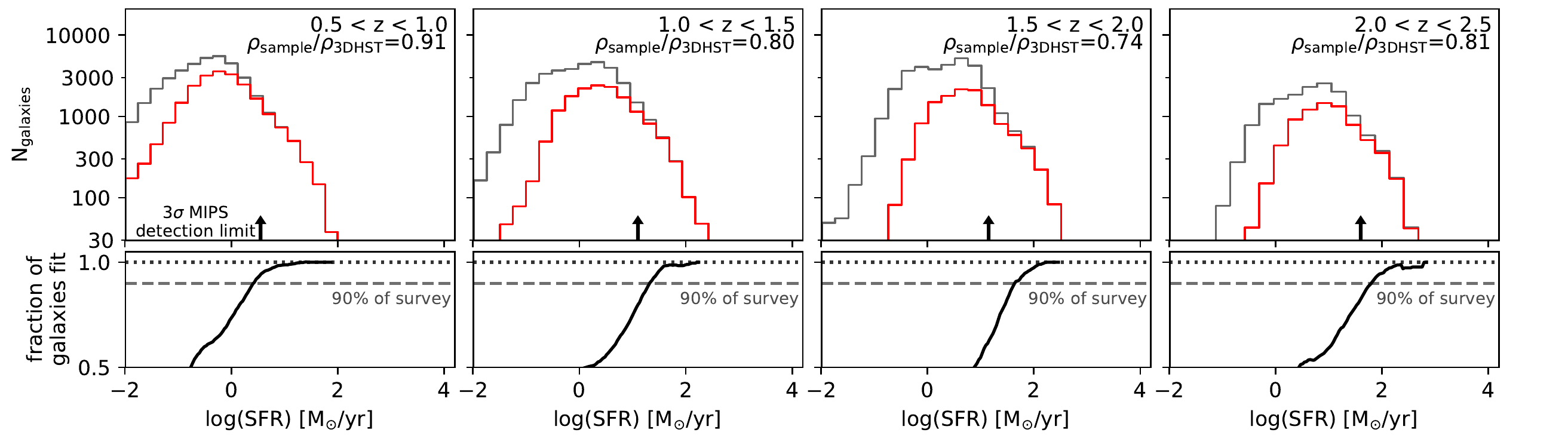}
\caption{Sample selection completeness in stellar mass and star formation rate. The black lines in the histograms represent the number of galaxies in the full 3D-HST photometric catalogs while the red lines represent the subset fit with the \mname{} model. Stellar masses come from FAST and the star formation rate shows \sfruvir{}. The fraction of the (stellar mass density/star formation rate density) measured by the target sample is indicated in the upper right of each panel, where the total is taken to be the full 3D-HST sample in that redshift range. The 95\% completeness limit is marked in grey.}
\label{fig:coverage}
\end{center}
\end{figure*}
There are 176,146 galaxies in v4.1 of the 3D-HST catalogs with usable photometry and derived stellar populations parameters from FAST \citep{kriek09}. Due to computational constraints we do not fit the entire sample in this work. Here we describe the subsample of galaxies which are fit. We also calculate the fraction of stellar mass density and star formation rate density (SFRD) covered by this sample in order to put measurements of the SFRD in Section \ref{sec:sfrd} in proper context.

We fit all galaxies above the FAST stellar mass completeness limit from \citet{tal14} between $0.5 < z < 2.5$ which have usable photometry (i.e., 3D-HST $use\_phot$ = 1). We include a small fraction of galaxies which are below the mass limits but have high-quality data according to the following criteria:
\begin{itemize}
\item S/N(F160W) $> 10$
\item $0.5 < z < 2.5$
\item $\sigma_\mathrm{z} < 0.25 $
\item 3D-HST $use\_phot$ = 1
\end{itemize}
These cuts result in 58,461 galaxies, of which 2702 (5\%) have measured z$_{\mathrm{spec}}$, 12,513 (21\%) use z$_{\mathrm{grism}}$, and the remaining 43,246 (74\%) use z$_{\mathrm{phot}}$. The target sample is $\sim$33\% of the total 3D-HST catalog by number, but covers the majority of the observed star formation rate density ($\gtrsim74\%$) and the stellar mass density ($\gtrsim95\%$) at $0.5 < z < 2.5$ (Figure \ref{fig:coverage}).

This subsample has the reliable photometry and high signal-to-noise in the detection bands where it is most efficient to fit the computationally intensive \mname{} model. The higher S/N data provide stronger parameter constraints. Additionally, the redshift quality cuts ensure that systematic errors due to redshift uncertainties are minimized (future prospects for propagating redshift uncertainties to the SED parameters are discussed in Section \ref{sec:zerr}). The galaxies removed by these cuts thus either have uncertain photometry, uncertain redshifts, or both.

The price of creating a computationally tractable sample is completeness: not every galaxy in the 3D-HST catalogs has an associated \prospector{} fit. The completeness of the target sample in FAST stellar mass and \sfruvir{} is shown in Figure \ref{fig:coverage}. Galaxies in the 3D-HST photometric catalog with $use\_phot$ = 1 are taken as the master sample against which this completeness is inferred. The fraction of the total stellar mass and total SFR covered by the target sample in each redshift window is indicated in the upper-right of each panel. 95-100\% of the total stellar mass and 74-91\% of the total star formation rate is covered by our target sample.

In some cases, the incompleteness due to imaging depth becomes comparable to the incompleteness due to the sub-sampling of the catalog. The 90\% completeness in FAST stellar mass is taken from \citet{tal14}, and are derived by comparing object detection rates in the CANDELS deep fields with a re-combined subset of the exposures which reach the depth of the CANDELS wide fields. The completeness in \sfruvir{} is taken as the 3$\sigma$ 24$\mu$m depth calculated in \citet{whitaker14} and represents where the observable constraint on IR star formation rates starts to become unreliable.

\subsection{Treatment of photometric zero points}
The 3D-HST team self-consistently re-derive zero points for each instrument and filter. This is necessary to bring data from many telescopes and instruments onto a common flux scale. This procedure is described in detail in the Appendix of \citet{skelton14}. In brief, every galaxy is fit by the photometric redshift code EAZY, and the systematic residuals between the EAZY templates and the observed photometry are tabulated. In general, the systematic residuals are caused by a combination of template mismatch and zero point errors. These two effects can be distinguished with sufficient quantities of high-quality data, as template mismatch occurs in the {\it rest} frame, while zero point errors are in the {\it observed} frame. The resulting derived zero point errors are used to correct the raw 3D-HST photometric fluxes to the fluxes reported in the catalog.

However, this process is imperfect: the `edges' of the wavelength coverage (IRAC 4 and $U$-band) are more poorly calibrated, and effects such as redshift-dependent template mismatch may also be folded into the derived zero point offsets. To avoid potentially imprinting any systematic offsets from this process into the photometry, we add the zero point correction for each band of photometry to the flux errors in quadrature. This effect varies from $0-28$\% of the total flux, depending on the photometric band. 

The $HST$ zero points are considerably more stable than the other bands, and are therefore treated differently. For HST bands the zero point corrections derived by the 3D-HST team are removed (these are typically near zero, though can be up to 8\% of the total flux), and no inflation of photometric errors is performed.

After this process, a 5\% minimum error is enforced for each band of photometry to allow for systematic errors in the physical models for stellar, gas, and dust emission.

\section{SED Modeling}
\label{sec:sedmodel}
\subsection{The \mname{} physical model}
\label{sec:physmodel}
\begin{figure*}[t!h!]
\begin{center}
\includegraphics[width=0.95\linewidth]{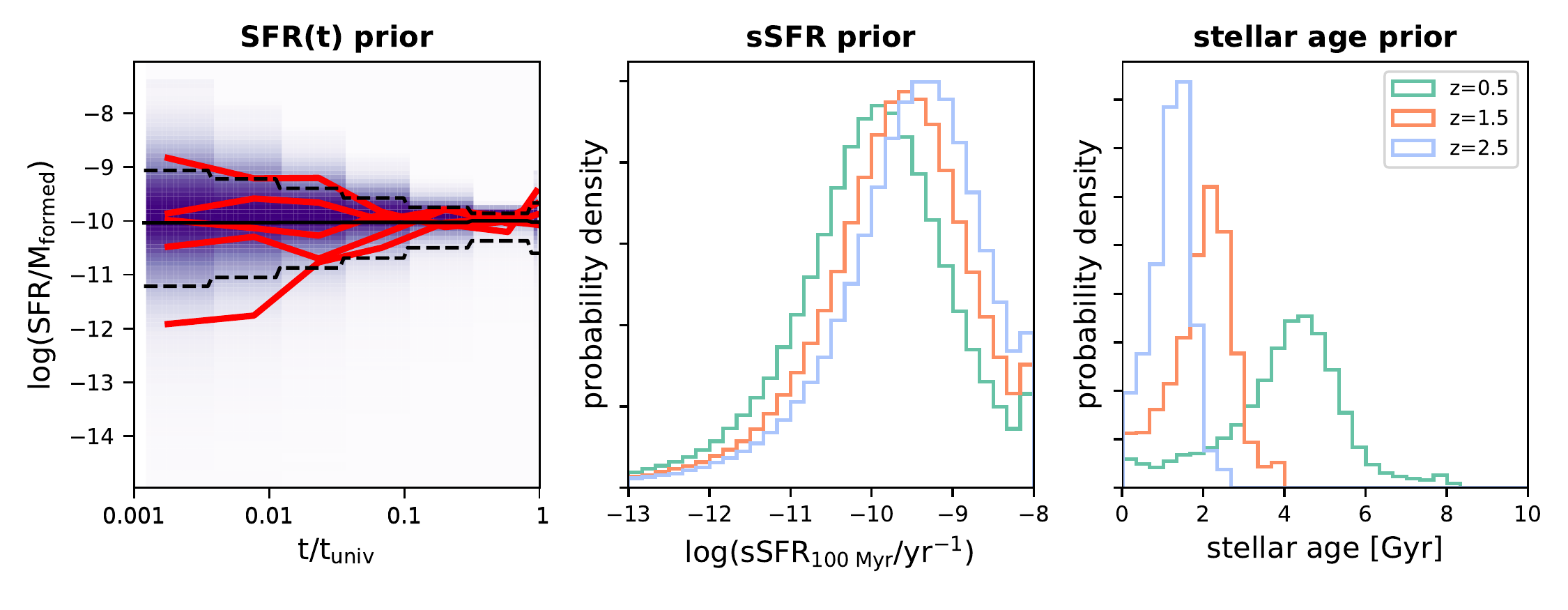}
\caption{The continuity SFH prior adopted in the \mname{} model. The left panel shows the prior density for SFR(t), with 5 random draws from the prior illustrated in red. The solid black line shows the median while the dashed black lines show the 16$^{\mathrm{th}}$ and 84$^{\mathrm{th}}$ percentiles. The middle and right panels show the prior in sSFR(100 Myr) and mass-weighted age as a function of redshift. See \citet{leja19} for further details.}
\label{fig:sfh_prior}
\end{center}
\end{figure*}

\begin{deluxetable*}{|p{0.1\textwidth}|p{0.4\textwidth}|p{0.4\textwidth}|}
\tablecaption{Free parameters and their associated priors for the \mname{} physical model. \label{table:priors}}
\tablehead{
\colhead{Parameter} & \colhead{Description} & \colhead{Prior}
}
\startdata
log(M/M$_{\odot}$) & total stellar mass formed & uniform: min=7, max=12.5\\
log(Z/Z$_{\odot}$) & stellar metallicity & clipped normal: min=-1.98, max=0.19, mean and $\sigma$ from the \citet{gallazzi05} mass--metallicity relationship (see Section \ref{sec:physmodel})\\
SFR ratios & ratio of the SFRs in adjacent bins of the 7-bin nonparametric SFH (6 parameters total) & Student's-t distribution with $\sigma=0.3$ and $\nu=2$. \\
\dusttwo{} & diffuse dust optical depth & clipped normal: min=0, max=4, mean=0.3, $\sigma$=1\\
\dustone{} & birth-cloud dust optical depth & clipped normal in (\dustone{}/\dusttwo{}): min=0, max=2, mean=1, $\sigma$=0.3 \\
\didx{} & power-law modifier to the shape of the \citet{calzetti00} attenuation curve & uniform: min=-1, max=0.4 \\
log(Z$_{\mathrm{gas}}$/Z$_{\odot}$) & gas-phase metallicity & uniform: min=-2, max=0.5 \\
\fagn{} & AGN luminosity as a fraction of the galaxy bolometric luminosity & log-uniform: min=10$^{-5}$, max=3 \\
\tauagn{} & optical depth of AGN torus dust & log-uniform: min=5, max=150\\
\enddata
\end{deluxetable*}

We use the \prospector{} inference framework \citep{leja17,prospector17} to construct a galaxy SED model. \prospector{} adopts a Bayesian approach to forward-modeling galaxy SEDs.

\prospector{} uses the Flexible Stellar Population Synthesis (\fsps{}) code \citep{conroy09b} to generate the underlying physical model and \pythonfsps{} \citep{pythonfsps14} to interface with \fsps{} in python. The physical model uses the MIST stellar evolutionary tracks and isochrones \citep{choi16,dotter16} based on MESA, an open-source stellar evolution package \citep{paxton11,paxton13,paxton15}. 

Notably, MIST models include stellar rotation. This has several salient effects on massive star evolution: (i) it channels additional fuel into the stellar core, (ii) it causes distortion in the shape of the star causing the poles to be hotter than the equator (`gravity darkening'), and (iii) it boosts the effect of mass loss. The net result is hotter, brighter, and longer-lived (by between 5-20\%) massive stars \citep{choi16}. This in turn causes higher UV and ionizing photon production in stellar populations between 0-20 Myr \citep{choi17}, especially at sub-solar metallicities. We note that models for rotation in stars are sensitive to implementation details and their predictions vary substantially; for example the Geneva rotation model \citep{ekstrom12} predicts both hotter and brighter stars than the MIST models adopted here (Fig. 1, \citealt{gossage18}). These models are currently difficult to test due to both the relative lack of nearby star clusters with populated main-sequence turnoffs and the ongoing debate about the similar effects of age spread and rotation \citep{bastian09,goudfrooij14,piatti17,gossage18}. While the physical mechanisms are distinct, the net effect on galaxy SEDs is similar to but weaker than the effect of stellar binaries \citep{eldridge17}.

In this study, we use an adapted version of the \mname{} model framework described in \citet{leja17,leja18}. The \mname{} model includes a nonparametric star formation history, a two-component dust attenuation model with a flexible attenuation curve, variable stellar metallicity, and dust emission powered via energy balance. Nebular line and continuum emission is generated self-consistently through use of CLOUDY \citep{ferland13} model grids from \citet{byler17}. Extensive calibration and testing of this model has been performed on local galaxies \citep{leja17,leja18}

We make multiple changes to the \mname{} model in order to reflect the different physics of galaxies at higher redshifts and to tailor the model more closely to the wavelength coverage and S/N of the 3D-HST photometry. The full set of priors and parameter ranges for the adjusted 14-parameter \mname{} model are shown in Table \ref{table:priors}. The salient changes are described below.

{\bf Nonparametric star formation history prior}: the continuity prior described in \citet{leja19} is taken as the prior for the nonparametric SFR(t). In brief, this prior weights against sharp transitions in SFR(t), similar to the regularization schemes from \citet{ocvirk06,tojeiro07}. The prior is tuned to allow similar transitions in SFR(t) to those of galaxies in the Illustris hydrodynamical simulations \citep{vogelsberger14a,vogelsberger14b,torrey14,diemer17}, though it is deliberately set to encompass broader behavior than is seen in these simulations. The resulting prior probability density for SFR(t), mass-weighted age, and sSFR$_{\mathrm{100 \; Myr}}$ is shown in Figure \ref{fig:sfh_prior}.

{\bf Spacing of the nonparametric star formation history bins}: Seven time bins are used in the nonparametric star formation history model. The bins are specified in lookback time. Two bins are fixed at $0-30$ Myr and $30-100$ Myr to capture variations in the recent star formation history of galaxies. A third bin is placed at (0.85\tuniv{} - \tuniv{}), where \tuniv{} is the age of the Universe at the observed redshift, to model a ``maximally old" population. The remaining four bins are spaced equally in logarithmic time between 100 Myr and 0.85\tuniv{}.

{\bf Stellar mass--stellar metallicity prior}: A single stellar metallicity is fit for all stars in a galaxy. A modified version of the stellar mass--stellar metallicity relationship from $z=0$ Sloan Digital Sky Survey (SDSS) data \citep{gallazzi05} is adopted as a prior. The relationship is modeled as a clipped normal distribution with limits of $-1.98 < \log(\mathrm{Z/Z}_{\odot})<0.19$ set by the range of the MIST stellar evolution tracks. The standard deviation is taken as the (84$^{\mathrm{th}}-16^{\mathrm{th}}$) percentile range from the \citet{gallazzi05} $z=0$ relationship, i.e. twice the observed standard deviation of the $z=0$ relationship. This wider relationship is adopted to allow potential redshift evolution in the stellar mass--stellar metallicity relationship.

{\bf A fixed IR SED}: The rest-frame mid-infrared is poorly sampled by the 3D-HST photometric catalog, as the reddest two filters are $Spitzer$/IRAC channel 4 (7.8$\mu$m) and $Spitzer$/MIPS 24$\mu$m. This results in poor constraints on the shape of the IR SED (rest-frame $\sim$4-1000$\mu$m). Accordingly, we fix the shape of the IR SED in \mname{} such that the $Spitzer$/MIPS 24$\mu$m to \lir{}($8-1000\mu$m) conversion approximates that of the log-average of the \citet{dale02} templates \citep{wuyts08}. \citet{wuyts11a} show that this luminosity-independent conversion produce \lir{} estimates which are in agreement with observed $Herschel$/PACS photometry, though with significant scatter. Additionally, this choice of IR SED follows \citet{whitaker14}, which facilitates direct comparisons with \sfruvir{} from the 3D-HST catalog. Hot dust emission powered by an AGN of variable strength is also permitted in the \mname{} model \citep{leja18}-- notably, this energy balance is performed separately from the rest of the IR SED, which is powered solely by stellar emission. Future potential for a more flexible IR SED model in \mname{} is discussed in Section \ref{sec:fixedirsed}.

{\bf Altered nebular physics}: Observations suggest that the gas in star-forming galaxies at higher redshifts experiences more extreme ionizing radiation fields and has metallicity abundances which may differ significantly from their stellar abundances \citep{shapley15,steidel16}. Accordingly, the ionization parameter for the nebular emission model is raised from $\log(\mathrm{U})=-2$ to $\log(\mathrm{U})=-1$, and gas-phase metallicity is decoupled from the stellar metallicity and allowed to vary between $-2 < \log(\mathrm{Z/Z}_{\odot}) < 0.5$. This is a nuisance parameter for the majority of galaxies as it typically is very poorly constrained by the photometry, though it can be important for very blue galaxies with high sSFRs \citep{cohn18}.

\subsection{Posterior sampling}
The posteriors are sampled with the dynamic nested sampling code \dynesty{} \citep{speagle19}\footnote{https://github.com/joshspeagle/dynesty}. Nested sampling has a number of desirable properties over standard Markov Chain Monte Carlo (MCMC) sampling, including well-defined stopping criteria, easier access to independent samples, more sophisticated treatment of multi-modal solutions, and simultaneous estimation of the Bayesian evidence. Additionally, dynamic nested sampling can be performed such that samples are targeted adaptively during the fit to better sample specific areas of the posterior. Finally, internal testing with \prospector{} shows that \dynesty{} requires $\sim$2x fewer model calls to produce similar posteriors to MCMC methods, which translates to a $\sim$50\% decrease in run-time. Each galaxy takes an average of $\sim25$ CPU-hours to converge for our 14-parameter model, resulting in $\sim$1.5 million CPU-hours\footnote{As a useful point of comparison, at the time of this writing 1.5 million CPU-hours costs approximately \$20,000 on Amazon Web Services. This is $\sim$40\% of the cost of one observing night on the Keck telescopes.} to analyze the whole sample.

Unless indicated otherwise, all reported parameters are the median of the marginalized posterior probability function, with 1$\sigma$ error bars reported as half of the 84$^{\mathrm{th}}$-16$^{\mathrm{th}}$ interquartile range. The \prospector{} parameter file for this version of the \mname{} model is available online\footnote{\url{https://github.com/jrleja/prospector_alpha/blob/master/parameter_files/td_delta_params.py}}.

\subsection{Benchmark models for SFR and stellar mass}
The next section compares the stellar masses and star formation rates derived from the \mname{} fits to the fiducial inferences from the 3D-HST catalogs. The key physical assumptions made in the 3D-HST derivations are repeated here for completeness.

Stellar masses in the 3D-HST catalogs are calculated with FAST \citep{kriek09}, a grid-based $\chi^2$ minimization code. \citet{bruzual03} (BC03) stellar population synthesis models are used with a \citet{chabrier03} IMF, fixed solar metallicity, exponentially declining star formation histories, and a single dust screen with a \citet{calzetti00} attenuation law. There is no nebular or dust emission; accordingly, regions of the SED with significant dust emission ($\lambda_{\mathrm{rest}} \gtrsim 3\mu$m) are heavily downweighted and $Spitzer$/MIPS 24$\mu$m photometry is not included in the fit. Only the best-fit parameters are reported. These are interchangeably called the 3D-HST catalog masses or the FAST masses in the text.

Star formation rates are calculated with the following relationship from \citet{bell05}:
\begin{equation}
\label{eqn:uvir_sfr}
\mathrm{SFR}\;  [M_{\odot} \; \mathrm{yr}^{-1}] = 1.09 \times 10^{-10} (L_{\mathrm{IR}}+2.2L_{\mathrm{UV}}) \; [L_{\odot}],
\end{equation}
with \lir{}($8-1000\mu$m) estimated directly from the $Spitzer$/MIPS 24$\mu$m flux and \luv{}($1216-3000\AA$) determined from the best-fit EAZY template \citep{whitaker14}. This conversion does not include any additional information about the composition of the underlying stellar populations. These are interchangeably called the 3D-HST catalog SFRs or \sfruvir{} in the text.
\begin{figure*}[t!]
\begin{center}
\includegraphics[width=0.95\linewidth]{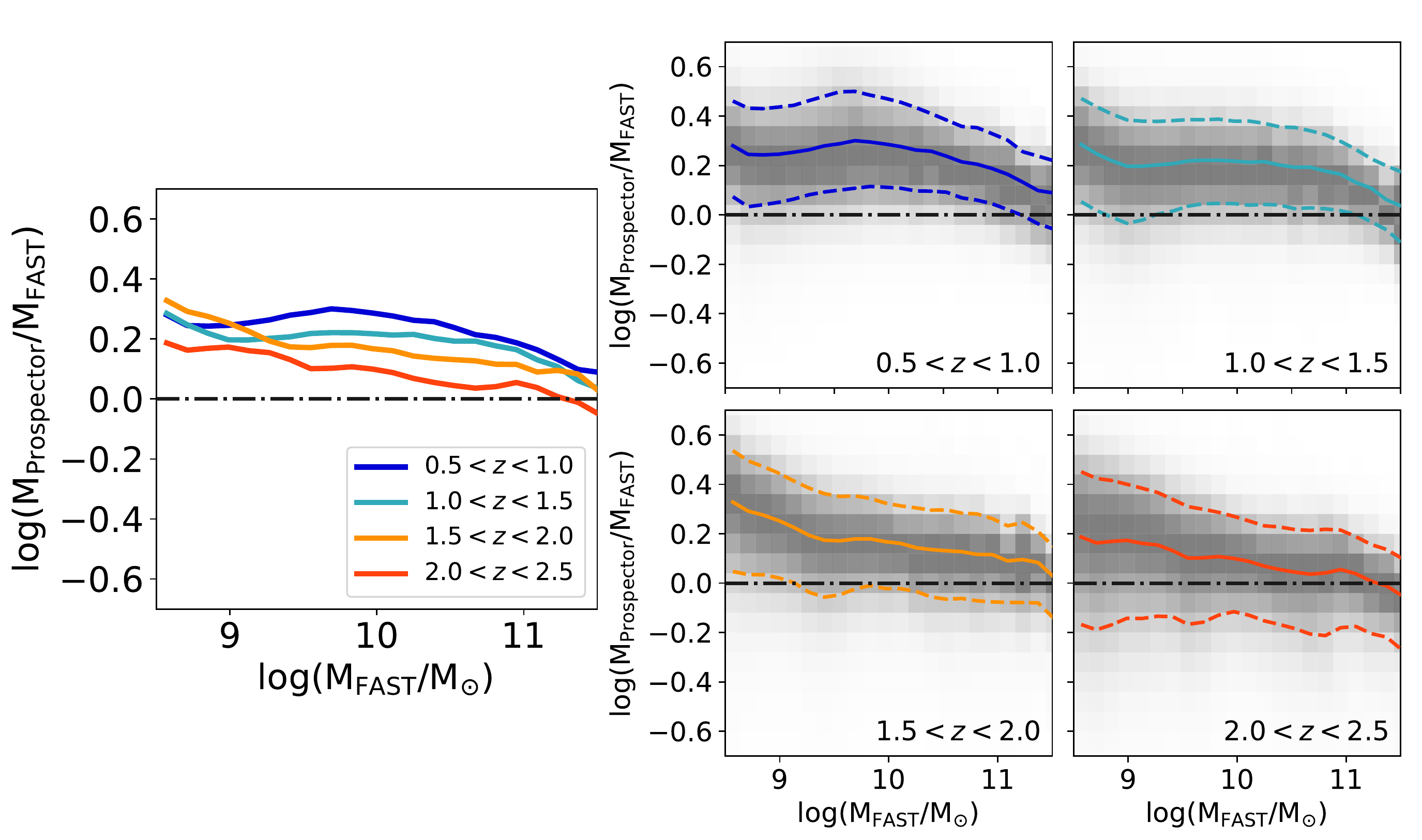}
\caption{\mname{} infers larger stellar masses than FAST. The right panels show the ratio of stellar masses in four redshift windows, while the left panel shows the median from each redshift window. The offset increases with decreasing redshift and increases with decreasing stellar mass. The grey shading is proportional to the stacked probability distribution function. The median is indicated by a colored solid line and the 16$^{\mathrm{th}}$ and 84$^{\mathrm{th}}$ percentiles are indicated by colored dashed lines.}
\label{fig:deltam_vs_z}
\end{center}
\end{figure*}

\section{Results}
\label{sec:deltased}
Stellar masses and star formation rates are among the most basic and important outputs of galaxy SED fitting codes and are therefore critical benchmarks for cross-code comparison. Here we compare the stellar masses and SFRs inferred from \mname{} to the fiducial masses and SFRs in the 3D-HST catalogs. There are systematic offsets in this comparison such that \mname{} masses are higher and the SFRs are lower. We demonstrate that the most significant causes of these offsets are older stellar populations and dust heating from old stars, respectively.

\subsection{Revised stellar masses}
\label{sec:mass_systematics}
Stellar mass is generally considered to be one of the most robust outputs of SED fitting, with typical systematic variations of $\sim$0.2 dex between codes (e.g., \citealt{mobasher15}). Though robust when compared to other outputs, systematic uncertainties of 0.2 dex in stellar masses result in critical uncertainties when interpreting dynamical masses, measuring galaxy mass assembly rates, and calibrating simulations of galaxy formation.

Figure \ref{fig:deltam_vs_z} shows the difference between the 3D-HST catalog masses and \prospector{} masses as a function of redshift. Specifically, the probability function for log(M$_{\mathrm{Prospector}}$/M$_{\mathrm{FAST}}$) as a function of log(M$_{\mathrm{FAST}}$) is created by summing the individual PDFs for all galaxies. The individual PDFs are calculated with the best-fit 3D-HST stellar masses and the full posterior distribution for the \mname{} stellar masses. As the 3D-HST stellar masses do not include error estimates, they are assigned a Gaussian PDF with a uniform width of 0.1 dex. The stacked PDFs thus include both galaxy-to-galaxy scatter and measurement uncertainty.

The correlation of the offset with mass and redshift give important clues as to the cause of the offsets. The median stellar mass difference is $\sim0.1-0.2$ dex ($\sim25-60\%$) with a $68^{\mathrm{th}}$ percentile range between $0.2-0.4$ dex. As stellar mass increases, the offset becomes smaller and the distribution becomes tighter. The offset also increases with decreasing redshift, with a larger increase at lower masses.

\begin{figure*}
\begin{center}
\includegraphics[width=0.95\linewidth]{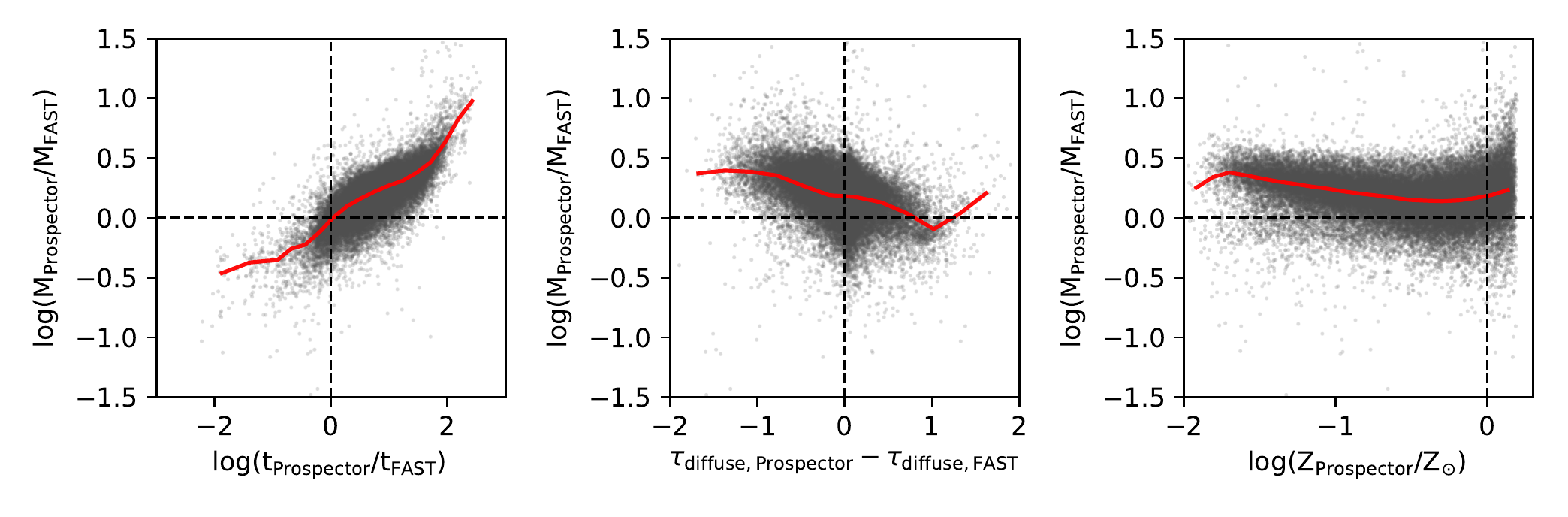}
\caption{Correlations between the difference in stellar mass and properties derived from SED fitting. From left to right, the x-axis shows the difference in mass-weighted age, the optical depth from diffuse dust attenuation, and stellar metallicity. The running median is highlighted in red. Stellar age appears to be most closely associated with the stellar mass difference between the 3D-HST catalog values and \mname{} and thus the likeliest cause of the offset.}
\label{fig:deltam_cause}
\end{center}
\end{figure*}

One potential cause of the mass offset is that FAST and \prospector{} use different stellar population synthesis codes (BC03 versus FSPS, respectively). The modularity of \prospector{} makes it possible to build a physical model in the \prospector{} framework which mimics the FAST physical model, thereby isolating the effect of different SPS codes. This FAST-like model is fit to a fraction of the 3D-HST catalog ($\sim$2700 galaxies) and the resulting mass offset is log(M$_{\mathrm{FSPS}}$/M$_{\mathrm{BC03}}) \approx 0.05$ dex. This implies that different stellar population synthesis codes contribute to, but do not dominate, the observed mass offset.

The bulk of the difference must then come from other differences in the SED models. Figure \ref{fig:deltam_cause} explores three primary candidates: the mass-weighted stellar age, the stellar metallicity, and the dust optical depth. The FAST mass-weighted stellar ages are calculated from the best-fit FAST SFH, while the \prospector{} mass-weighted ages are calculated from samples of the SFH posterior. As the stellar metallicity is fixed to solar in the 3D-HST catalog fits, the variable \mname{} metallicity is shown alone. The dust attenuation models have substantial differences: here we compare the V-band dust optical depth from the 3D-HST catalogs (computed with a fixed \citealt{calzetti00} attenuation curve) to the \mname{} V-and diffuse dust optical depth (computed with a flexible attenuation curve), which is only one component of the two-component \citet{charlot00} dust model in \mname{}. The relative difference in the V-band dust optical depths is a good proxy for the differential attenuation between each model.

Figure \ref{fig:deltam_cause} makes it clear that, of the model differences considered, the age differences are the primary driver of the systematic offset in stellar mass. Indeed, older stellar ages provide a clean explanation for the trend in median mass offset with redshift and stellar mass. The trend with redshift comes from the dependence of age on t$_{\mathrm{universe}}$(z): as redshift decreases, the upper limit on stellar age increases. This results in larger relative age differences permitted between \mname{} and the 3D-HST catalog inferences. The offset increases with decreasing stellar mass because low-mass galaxies are primarily blue and star-forming: these galaxies display the most sensitivity to the SFH parameterization and priors \citep{leja19}.

Notably, the systematic mass differences suggest that \mname{} will modify or break the tight relationship between mass-to-light (M/L) and optical color \citep{bell01}. As may be expected, \mname{} finds an increased M/L ratio at fixed optical color. It also finds greatly increased scatter in this relationship. This can broadly be attributed to the fact that a more complex physical model allows a wider range of physical properties at fixed optical color. This scatter in M/L is associated with variations in stellar age, metallicity, and the shape of the dust attenuation curve, and will be explored further in future work.

\subsection{Contrasting pictures of galaxy star formation histories}
The previous section demonstrated that differences in galaxy star formation histories can cause systematics in inferred stellar masses. These differences in SFR(t) can be substantial: the mass-weighted ages inferred in the 3D-HST catalog and \mname{} differ by factors of 3-5 for the majority of the galaxy population, despite being constrained by the exact same photometry. There are several reasons that SFHs are typically only weakly constrained by broadband photometry:
\begin{enumerate}
\item Younger stars (t $\lesssim100$ Myr) dominate the observed SEDs of star-forming galaxies, greatly out-shining older stars \citep{maraston10},
\item Stellar isochrones evolve very little at late ages (t $\gtrsim2$ Gyr), making it relatively difficult to distinguish between different age models for older galaxies \citep{conroy13a},
\item Stellar age, stellar metallicity, and dust have similar effects on the UV-NIR SED which can result in significant parameter degeneracies \citep{bell01}.
\end{enumerate}
When the data provide poor constraints, the prior for SFR(t) becomes very important in determining the output \citep{carnall19,leja19}. The prior on SFR(t) is determined both by the chosen SFH parameterization and by the priors on each parameter. Crucially, sensitivity to the prior is not specific to Bayesian analysis; classical methods implicitly set a uniform prior over the chosen SFH parameterization and range of the parameter grids. The continuity prior in \mname{} is qualitatively very different than the exponentially declining SFH assumed in the 3D-HST analysis, so the difference in recovered SFHs is not surprising. 

\begin{figure*}
\begin{center}
\includegraphics[width=0.95\linewidth]{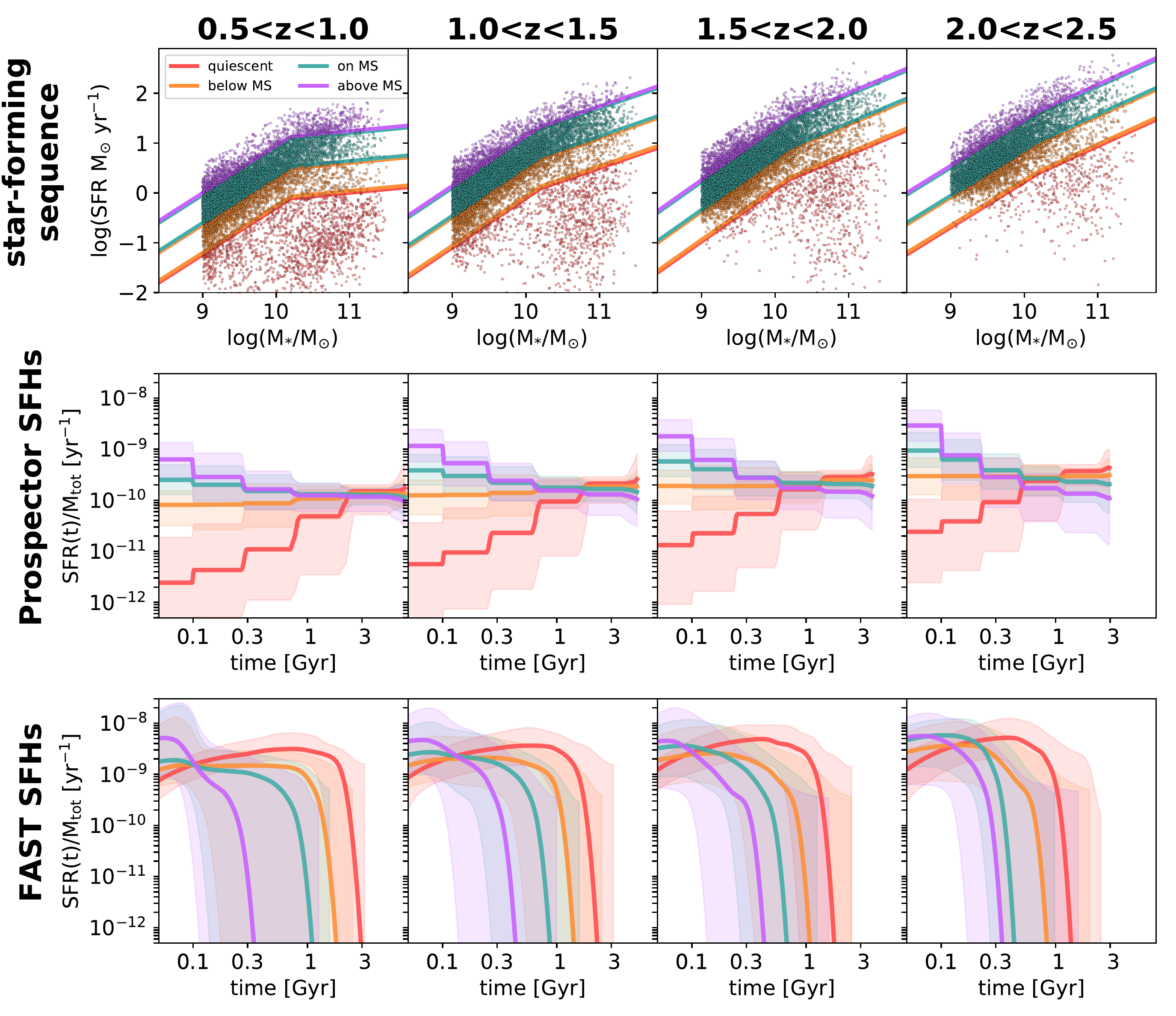}
\caption{Stacked star formation histories from \mname{} and FAST as a function of star formation rate, stellar mass, and redshift. The upper row of panels show the distribution of galaxies in the star-forming sequence. Galaxies are divided into above, on, and below the star-forming sequence, and quiescent and their SFHs are stacked separately. The two lower rows of panels show the median of the SFH stacks and the shaded regions cover the 16$^{\mathrm{th}}$ and 84$^{\mathrm{th}}$ percentiles from both \mname{} and the 3D-HST catalogs. The 3D-HST catalog SFHs produce stellar populations which are far younger (factors of 3-5 and more) than the \mname{} SFHs.}
\label{fig:ms_stack}
\end{center}
\end{figure*}

The SFHs inferred from these two analyses imply very different pictures of galaxy evolution. Figure \ref{fig:ms_stack} shows star formation histories stacked across the star-forming sequence from both the 3D-HST analysis and the \mname{} fits. These stacks are comprised of galaxies split into four categories: above, on, and below the star-forming sequence, and quiescent. For consistency, star formation rates from \mname{} are used to sort galaxies in both stacks (the FAST SFRs are unreliable as they do not include IR constraints). The locus of the star-forming main sequence is taken from \citet{whitaker14} and corrected downwards by 0.3 dex to account for the typical difference between SFR$_{\mathrm{Prospector}}$ and \sfruvir{} (see Section \ref{sec:sfr}). The vertical divisions are taken to be 0.6 dex wide, or roughly twice the logarithmic scatter in the main sequence \citep{speagle14}. The SFH stacks are created by summing the individual PDFs for SFR(t)/M$_{\mathrm{formed}}$\footnote{Note that sSFR is calculated using {\it stellar} mass but SFR(t) is normalized by {\it total mass formed}. This causes some overlap in the youngest star formation history bins, which would be strictly forbidden if the definitions of mass were the same.} such that each galaxy in the stack is weighted equally. 

The most striking result in Figure \ref{fig:ms_stack} is the contrast in average galaxy age. For example, the FAST fits infer that at 0.5 $< z <$ 1, galaxies above the star-forming sequence are $\sim 200-300$ Myr old while galaxies on the star-forming sequence are $\sim$ 1 Gyr old. In contrast, the \mname{} SFHs infer galaxy ages of order a few Gyr regardless of their position on the star-forming sequence. These SFHs imply very different galaxy mass assembly histories. We demonstrate via a continuity analysis (Section \ref{sec:continuity}) that the assembly histories implied by the 3D-HST fits are far too rapid to be consistent with the observed evolution of the stellar mass function.

There are also strikingly different descriptions of a galaxy's lifetime on the star-forming sequence. The \mname{} SFHs find that galaxy ages show little correlation with their position relative to the star-forming sequence. Indeed, the \mname{} SFHs are consistent with a galaxy's position on the star-forming sequence being a temporary status, lasting of order $\sim100-500$ Myr before converging on long-term SFHs with similar trajectories. On the other hand, the 3D-HST fits imply that a galaxy's position relative to the star-forming sequence is strongly correlated with its lifetime, with galaxies above the main sequence having appeared between $300-500$ Myr in the past and galaxies on the star-forming sequence having lifetimes of $\sim$ 1 Gyr. This is almost a necessary conclusion when fitting exponentially declining SFHs, as the only way to generate relatively high sSFRs in such a framework is to have very young ages. 

A common rationale for using exponentially declining SFHs is that the inferred $\tau$ and age are meant to characterize the bulk of the most recent star formation rather than representing an actual SFH. However, given that the actual SFH implied by these models directly affects the mass estimate, it is more useful in this comparison to take the SFHs at face value.

Beyond the cross-comparison, the \mname{} SFHs in Figure \ref{fig:ms_stack} provide an interesting overview of galaxy formation and evolution over the critical period of $0.5 < z < 2.5$. The \mname{} stacks show that at higher redshifts, typical galaxies on and above the star-forming sequence have rising SFHs while those below the star-forming sequence have flatter SFHs. Galaxies above the star-forming sequence at $2 < z < 2.5$ were on this sequence $\sim$ 100 Myr in the past, while galaxies below the star-forming sequence have been off this sequence for three times longer. Quenched galaxies have falling SFHs and get older with decreasing redshift, and have also been quenched for longer at lower redshifts.

Given the dominant role of the prior in the recovery of star formation histories \citep{carnall18,leja18}, it will be important to establish which of the \mname{} results are driven by the data and which are driven by the prior. For example, it is unlikely that the data are constraining the characteristic timescales on which star formation rates change as a broad range of characteristic timescales are often equally consistent with constraints from broadband photometry \citep{leja18}. A more detailed analysis of these trends is deferred to future work.

\begin{figure*}[ht!]
\begin{center}
\includegraphics[width=0.95\linewidth]{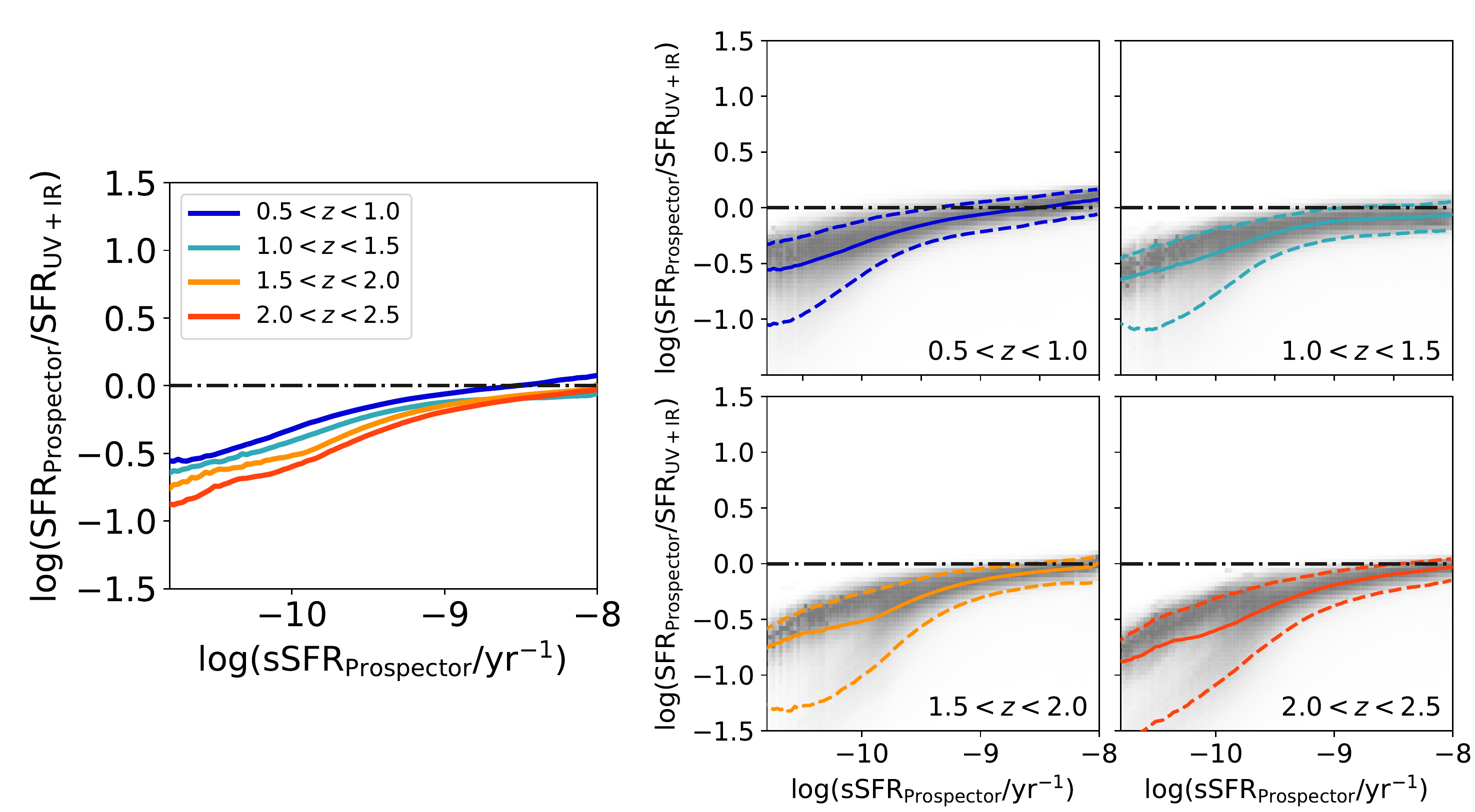}
\caption{The offset between \sfruvir{} and SFR$_{\mathrm{Prospector}}$ as a function of sSFR$_{\mathrm{Prospector}}$. The right panels show four different redshift windows with grey shading representing the stacked probability distribution function. The median is a colored solid line and the 16$^{\mathrm{th}}$ and 84$^{\mathrm{th}}$ percentiles are colored dashed lines. The left panel highlights the redshift evolution of the median. There is good agreement at high sSFR, but at lower sSFRs the \mname{} SFRs are increasingly lower than \sfruvir{}.}
\label{fig:ssfr_uvir_comparison}
\end{center}
\end{figure*}

\subsection{Revised star formation rates}
\label{sec:sfr}
UV+IR star formation rates are considered more reliable than those from SED fitting codes such as FAST because they also include contributions from dusty star formation via the observed IR luminosities. However, these values do not include galaxy-to-galaxy variation in the underlying stellar populations properties which is measured directly in SED fitting. Here we show that SFRs from panchromatic SED fitting are systematically lower than \sfruvir{}, and that this offset is largely due to energy emitted from older stellar populations. This includes energy observed directly in the UV and energy attenuated and re-emitted by dust.

The 3D-HST catalogs provide \sfruvir{} from Equation \ref{eqn:uvir_sfr} following the methodology of \citet{whitaker14}. \lir{} is obtained in the 3D-HST analysis by converting the observed $Spitzer$/MIPS 24$\mu$m flux directly into \lir{} using a fixed template. However, the observed IR fluxes are not reliable for low-mass galaxies due to confusion limits. To extend this comparison to low-mass galaxies, we instead calculate the $Spitzer$/MIPS 24$\mu$m flux from model spectra drawn from the \mname{} posteriors. These are combined with the log-average of the \citet{dale02} templates to calculate \lir{}. \luv{} is measured directly from the \mname{} model spectra. 

To ensure that the resulting SFR$_{\mathrm{UV+IR}}$ values are not systematically biased by this approximation, we compare UV+IR SFRs calculated from the posteriors of the \mname{} model fits to the UV+IR SFRs from \citet{whitaker14}. There is no measurable offset as a function of SFR and there is a relatively low scatter of 0.24 dex, suggesting the model \sfruvir{} are an acceptable approximation for the values in the 3D-HST catalog.

Figure \ref{fig:ssfr_uvir_comparison} shows the stacked distribution of \sfruvir{} / SFR$_{\mathrm{Prospector}}$ as a function of sSFR$_{\mathrm{Prospector}}$. This is created by summing the individual probability distribution functions for all galaxies. The median offset ranges between $0-1$ dex and is largest at low sSFRs. The central 68$^{\mathrm{th}}$ percentile ranges from $0.2-0.8$ dex and is also largest at low sSFRs.

\begin{figure*}
\begin{center}
\includegraphics[width=0.95\linewidth]{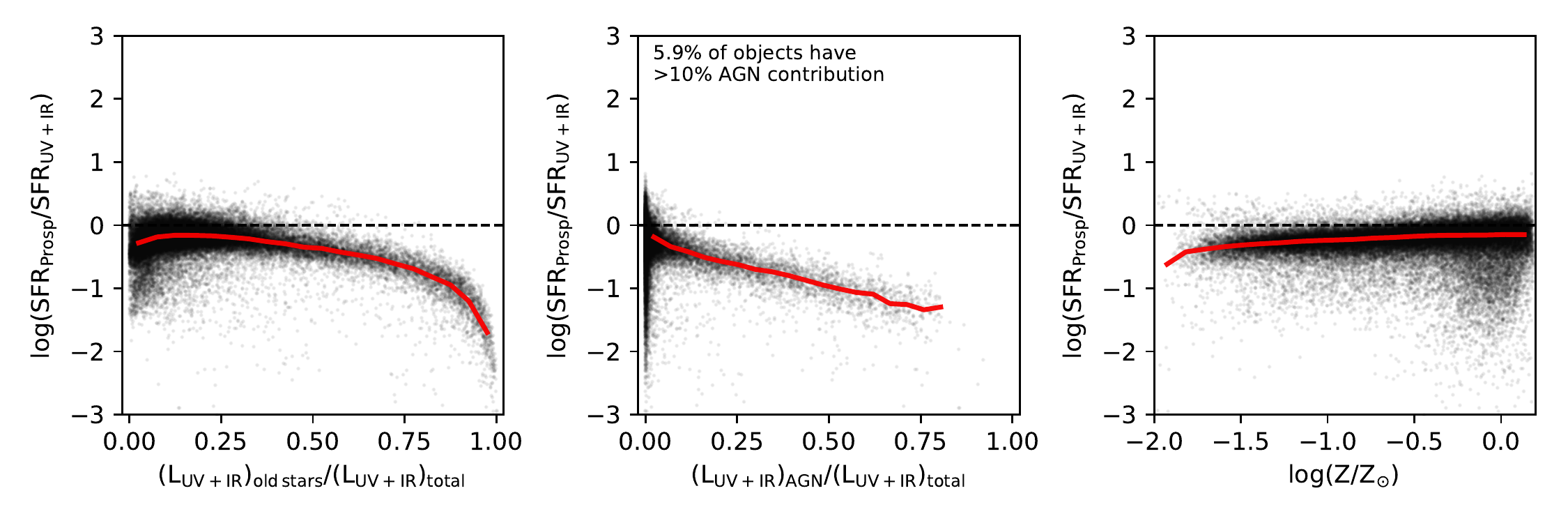}
\caption{Correlations between SFR$_{\mathrm{Prospector}}$/\sfruvir{} and derived galaxy properties. From left to right, the x-axis values are the fraction of \lir{}+\luv{} emitted by old stars, the fraction of \lir{}+\luv{} emitted by AGN, and the stellar metallicity. While all three components are correlated with the offset, the offset correlates most clearly with heating by old stellar populations ('old' defined here as $t>100$ Myr). We note that while AGN are strongly associated with lower star formation rates, they affect a relatively small proportion of the population (only 5.9\% of galaxies have an AGN contribution of $>10\%$).}
\label{fig:deltasfr_cause}
\end{center}
\end{figure*}

Figure \ref{fig:deltasfr_cause} explores potential physical causes of this offset: additional flux from ``old'' ($t>$100 Myr) stellar populations, hot dust emission from AGN activity, and a nonsolar stellar metallicity. The x-axis of the left two panels shows the fractional change in (\luv{} + \lir{}) when old stars and AGN are removed from the \mname{} model, while the third panel simply shows log(Z/Z$_{\odot}$).

The offsets show some correlation with all three parameters, suggesting that the overall change in inferred SFR cannot be simply associated with a single cause. However, the clearest correlation is with energy from old stars. This effect naturally explains the trend of increasing offset with decreasing sSFR: at lower sSFRs, a higher fractional contribution of total flux is emitted by old stars. This energy from old stars includes both energy emitted directly in the UV and energy which is attenuated from the UV, optical, and near-infrared and re-emitted in the IR. Emission from buried AGN also strongly affect the star formation rate of a small fraction of galaxies, while stellar metallicity has a more subtle effect for many galaxies below $\mathrm{Z}=\mathrm{Z}_{\odot}$.

\subsection{Effect of old stellar heating on SFR estimates}
\label{sec:oldem}
Flux from old stars can have a strong effect on star formation rates inferred only from \luv{}+\lir{}. It is therefore important to clarify both how the strength of this effect varies across the galaxy population and how robustly this effect can be modeled within \prospector{}.

Equation \ref{eqn:uvir_sfr} for \sfruvir{} was derived by creating a stellar population with a constant SFR over 100 Myr. The underlying principle is energy balance: if all the observed luminosity comes from young stars, inverting this will return the number of young stars (i.e. the star formation rate). This is a good assumption when young stars dominate the stellar energy budget. However, old stars ($t > 100$ Myr) also contribute to the observed UV emission and indirectly to the observed IR emission via dust attenuation. This heating is undoubtedly occurring at some level: the salient question is to what extent it is important in affecting the simple \sfruvir{} estimates.

\begin{figure}
\begin{center}
\includegraphics[width=0.95\linewidth]{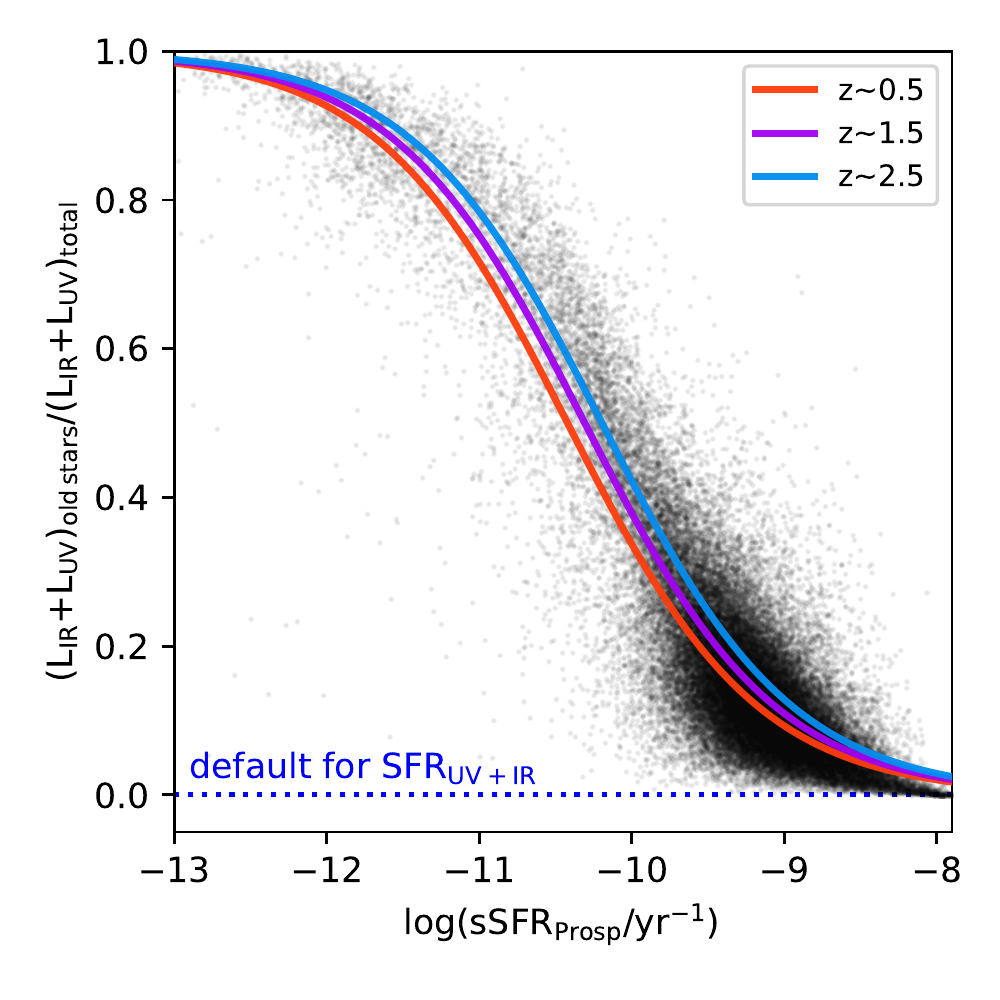}
\caption{Relationship between the fraction of \luv{}+\lir{} emitted by old stars ($t > 100$ Myr) and the sSFR inferred from panchromatic SED modeling. The fit to this relationship from Equation \ref{eqn:heating} is shown in red, while the $16^{\mathrm{th}} - 84^{\mathrm{th}}$ percentile range is shaded in red. As the specific star formation rate decreases, more and more of the luminosity is emitted by old stars. A linear transformation between UV+IR luminosity and star formation rate can thus overestimate the star formation rate for galaxies with low sSFR.}
\label{fig:heating_fraction_vs_ssfr}
\end{center}
\end{figure}

Figure \ref{fig:heating_fraction_vs_ssfr} shows the fraction of \luv{}+\lir{} emission originating from stars older than 100 Myr in the \mname{} model as a function of sSFR. The effect of old stellar heating on SFR estimates has been demonstrated at both low and high redshift for small samples \citep{cortese08,delooze14,utomo14} but the measurement presented here is the first for a statistically significant sample of galaxies. The relationship in Figure \ref{fig:heating_fraction_vs_ssfr} is fit with the equation
\begin{equation}
\label{eqn:heating}
y = 0.5\tanh\left(a\log\left[\mathrm{sSFR/yr}^{-1}\right]+bz+c\right)+1
\end{equation}
where $y = (L_{\mathrm{UV+IR}})_{\mathrm{old\;stars}}/(L_{\mathrm{UV+IR}})_{\mathrm{total}}$, $a=-0.8$, $b=0.09$, and $c=-8.4$.

As might be expected, galaxies with high sSFRs ($\gtrsim 10^{-9}$ yr$^{-1}$) experience negligible contribution from old stars, while galaxies with low specific star formation rates $\lesssim$ 10$^{10.5}$ yr$^{-1}$ are dominated by emission from old stars. The point of equal contribution is at sSFR $\approx$ 10$^{-10.3}$ yr$^{-1}$. For reference, a 10$^{10.5}$ M$_{\odot}$ galaxy on the star-forming sequence at $z=0.75$ has a specific star formation rate of $\sim$10$^{-9.4}$ yr$^{-1}$ \citep{whitaker14} and approximately 20\% of the observed IR and UV luminosity in such a galaxy is expected to come from old stars. This effect decreases to $<10\%$ at $z=2.25$.

\begin{figure*}
\begin{center}
\includegraphics[width=0.95\linewidth]{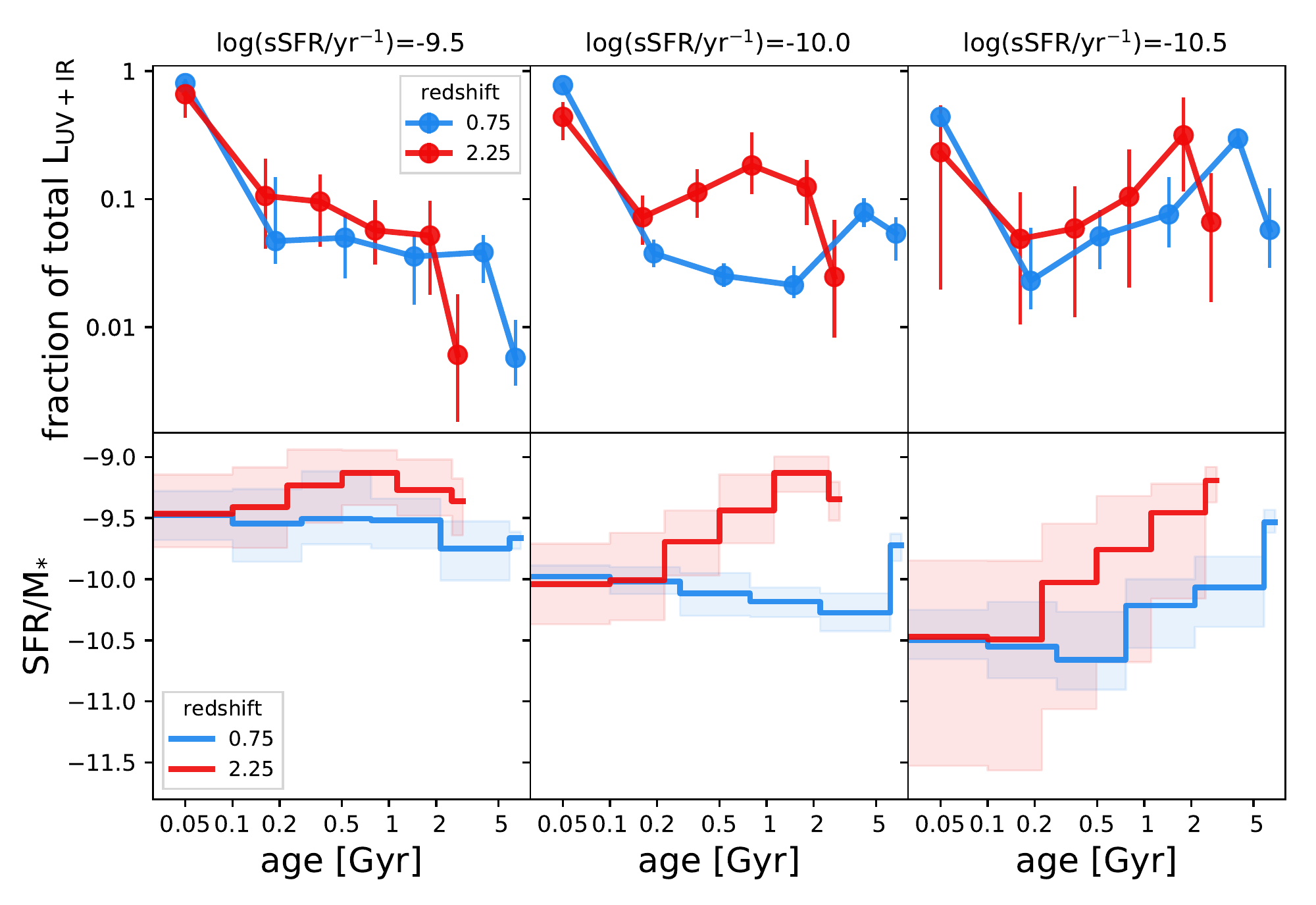}
\caption{The bulk of `old' stellar heating is typically performed by stars aged $0.1-1$ Gyr at $z=2.25$ and by stars aged $2-6$ Gyr at $z=0.75$. The upper panels show the relative contribution to L$_{\mathrm{UV}}$+L$_{\mathrm{IR}}$ from stellar populations of different ages, with the youngest bin encompassing 0$-$100 Myr. Each panel compares randomly selected galaxies from different redshifts at a fixed log(sSFR). In each case, the relative contribution of young ($< 100$ Myr) stars decreases with increasing redshift. The lower panels show the corresponding SFHs, highlighting the fact that the `old' stars in high-redshift galaxies are comparatively much younger and more luminous. This means that the strength of old stellar heating {\it increases} with redshift. Note that the consistent `dip' in the upper panels for the contribution is caused by the SFH bin spacing. The oldest time bin is approximately four times smaller than the previous bin, meaning there is consistently less mass in the oldest time bin.}
\label{fig:dust_heating_example}
\end{center}
\end{figure*}

Somewhat counterintuitively, the offset between SFR$_{\mathrm{Prospector}}$ and \sfruvir{} increases with increasing redshift (Figure \ref{fig:ssfr_uvir_comparison}), implying that old stars make up a larger fraction of the observed L$_{\mathrm{UV+IR}}$ in higher redshift galaxies. Fig. \ref{fig:dust_heating_example} confirms this, showing the fractional contribution to the total L$_{\mathrm{UV+IR}}$ from stars as a function of age and galaxy redshift. The bulk of `old' stellar heating is performed by stars aged $0.1-1$ Gyr at $z=2.25$ and by stars aged $2-6$ Gyr at $z=0.75$. The old stellar populations in high-redshift galaxies are comparatively younger and brighter, contributing more to L$_{\mathrm{UV+IR}}$ at a fixed value of sSFR. The strength of old stellar heating thus increases with redshift because the old stellar populations at $z=2$ are on average more luminous.

There is a good reason that this effect isn't typically included in SFR estimates: it is technically challenging to include the effect of dust heating from old stars as it requires that SFR, SFH, and dust attenuation be estimated from a single self-consistent model. In theory, it is possible to modify the assumed star formation history assumed in calculating \sfruvir{} to include more emission from old stars and reduce this bias \citep{kennicutt12}. This is not a universal solution though, as revising the recipe for \sfruvir{} in this fashion will then necessarily underestimate star formation rates in high sSFR galaxies.

Using a sophisticated model such as \prospector{} to estimate SFRs is not necessarily a panacea either. The fractional amount of energy generated by old stars depends not only on accurate estimates of the long-term SFH, but also on the spatial distribution of old and young stars relative to the dust. Thus the size of the effect in Figure \ref{fig:heating_fraction_vs_ssfr} is dependent on the adopted dust model. \mname{} uses a two-component \citet{charlot00} model wherein all stars are attenuated equally by a diffuse screen of dust, while younger stars experience extra attenuation. The variable shape of the dust attenuation curve adds more variance to the age-dependent attenuation, as wavelength-dependent attenuation translates into age-dependent attenuation due to the different emission profiles of young and old stars.

Assumptions about the star-dust geometry can be explored by observing systems where the contribution of old stars to the integrated UV and IR emission of galaxies can be separated. For example, the bulge of Andromeda is composed almost entirely of old stars and comprises 30\% of the total stellar mass yet only contributes 5\% of the IR luminosity. This may not be surprising, given the bulge also contains only 0.5\% of the total dust mass \citep{groves12b}. The majority of the dust lives in star-forming regions in the spiral arms. The key question, then, is to what extent the IR emission from the dusty spiral arms is caused by old stars, both nearby and from $\sim$kpc distances. This can be answered by spatially resolved modeling of mixed systems of old and young stars with a careful accounting of energy transfer between adjacent pixels. Studies which employ this approach find that a large fraction of the energy absorbed by dust in nearby spiral galaxies originates from the old stellar populations (e.g., 37\% for M51, 91\% for M31) \citep{delooze14,viaene17}.

Spatially resolved modeling may also have the potential to yield relationships which can better calibrate the energy contribution of old stars in unresolved SED modeling. For example, it is well-established that the dust temperature is closely related to the stellar density (e.g., \citealt{chanial07,rujopakarn11}), a relationship driven by the underlying relationship between dust temperature and the intensity of incident radiation. This means that systems which have different spatial distributions of young and old stars will show a wavelength-dependent infrared contribution from old stars. For example, direct $Herschel$ observations of Andromeda show that optical light from old bulge stars heat dust to higher temperatures than star-forming regions do \citep{groves12b}. Panchromatic radiative transfer models of Andromeda corroborate this picture, suggesting that dust heated only by old stars would peak at 150$\mu$m whereas younger stellar populations would cause it to peak around $200-250\mu$m \citep{viaene17}. The findings in Andromeda is generalizable to the nearby galaxy population: the KINGFISH survey \citep{skibba11} finds that, at fixed sSFR (i.e., a fixed ratio of young to old stars), early-type galaxies have hotter dust temperatures on average than late-type galaxies. This means that old stars are heating the dust to higher temperatures because radiation density is extremely high in dense stellar regions such as bulges where old stars happen to live. 

However, this relationship between dust temperature and stellar age can work in either direction depending on the relationship between stellar morphology and stellar age. Thus, galaxies which do not have a classic bulge-and-disk stellar morphology will likely behave differently: for example, \citet{chanial07} show that the overall dust temperature in galaxies is more closely related to the density of {\it young} stars. This relationship between young stars and hot dust emission is likely to become more dominant in shaping the overall SED of the galaxy at high redshift as stellar populations become younger (e.g., \citealt{imara18}). The relationship between stellar morphology, stellar age, redshift, and dust temperature is rich and complex, and deserves much deeper exploration in future work.

\section{Global implications and model cross-validation}
\label{sec:cross_validation}
The \mname{} model finds that on average, galaxies in the distant Universe are both more massive and more quiescent than suggested in previous studies. These effects are due to \mname{} inferring older ages and including the effect of old stellar heating respectively. In this section we examine the implications and the self-consistency of these results by cross-comparing different inferences of global quantities including the evolution of the stellar mass function and the cosmic star formation rate density. We also indirectly test the accuracy of \mname{} masses by comparing stellar and dynamical masses.
\subsection{The consistency between star formation histories and the growth of the stellar mass function}
\label{sec:continuity}
SED fitting simultaneously infers both the current stellar mass $M_*(t=0)$ and the past star formation history, d$M$/d$t(t)$. In principle this means that the galaxy stellar mass function $\phi(M,z)$ needs only to be observed at $z=0$; the redshift dependence of this function can then be predicted by evolving each galaxy backwards in time according to d$M$/d$t(t)$ while also accounting for the effect of galaxy mergers. In practice, the current stellar mass is a much more robust quantity than the SFH and so the mass function is better constrained by measuring the current stellar mass for galaxy populations across a range of redshifts. This ``redundant'' measurement creates an opportunity to test the self-consistency of SED fitting models. The inferred SFHs can be used to evolve the observed stellar mass function at a lower redshift $z_{\mathrm{start}}$ to some higher redshift $z_{\mathrm{obs}}$ and then compared with the observed stellar mass function at that redshift.

Here we perform this consistency check for the SFHs from \mname{} and from the FAST fits in the 3D-HST catalogs. To do this properly, it would be necessary to couple a \prospector{}-derived stellar mass function with a full \prospector{} analysis of SFHs and SFRs derived from different datasets at redshifts. As this full analysis has not yet been performed, here we instead recast the \prospector{} growth rates in terms of FAST mass functions measured in previous studies. This exercise will give a sense of what the full analysis might reveal, and indeed it looks promising.

We take the observed mass functions from \citet{tomczak14}, specifically adopting the smooth parameterizations of this mass function as a function of redshift from \citet{leja15} to ensure a monotonic evolution with redshift. The SED fitting in \citet{tomczak14} is performed using FAST, the same code used to generate the SED fitting outputs in the 3D-HST catalog, which ensures that there is minimal systematic offset between the mass function masses and the 3D-HST catalog masses. Accordingly, for consistency, the \prospector{} growth rate function is also cast in terms of the 3D-HST catalog mass.

For three initial redshifts z=(0.6, 1.1, 1.6), we select galaxies in a narrow range $\delta z=0.1$ and transform their SFHs into the distribution of fractional change in total mass formed $\Delta M_{\mathrm{formed}}/M_{\mathrm{formed}}$, hereafter called the growth kernel $f_M(z,M_*)$. For the \prospector{} results the kernel is built by summing the full PDFs; the 3D-HST results lack error estimates so the kernel is composed of the distribution of best-fit SFHs. The growth kernel $f_M(z,M_{*})$ is then smoothed in the mass direction, equivalent to assuming a smooth growth rate as a function of mass. Finally the mass function at a higher redshift $z_{\mathrm{obs}}$ is predicted by convolving the mass function observed at $z_{\mathrm{start}}$ by the growth kernel $f_M(z_{\mathrm{start}},M_{*}$).

We additionally include a simple model for the effect of galaxy-galaxy mergers on the stellar mass function from \citet{leja15}. In brief, this model includes effects from both the rate at which galaxies merge with more massive galaxies than themselves (i.e. the ``destruction'' rate) and the rate at which galaxies gain stellar mass from mergers (the ``growth'' rate) as a function of both stellar mass and redshift from the \citet{guo13b} semi-analytical model of galaxy formation. For this work we first increase the number density according to the destruction rate integrated between $z_{\mathrm{start}}$ and $z_{\mathrm{obs}}$, and then remove mass from galaxies according to the growth rate as a function of mass.

\begin{sidewaysfigure*}
\includegraphics[width=\textheight]{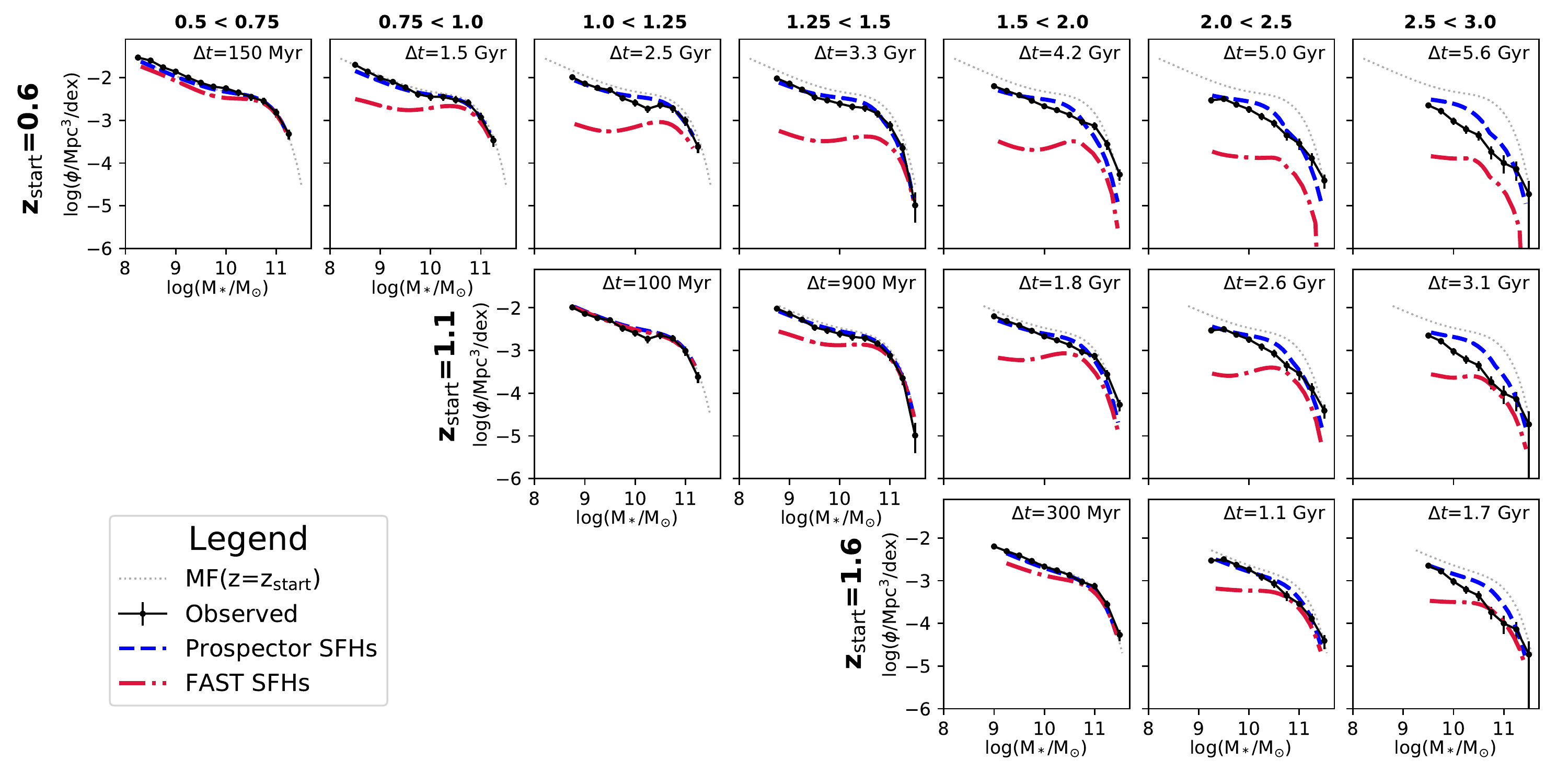}
\caption{Evolving the observed mass functions backwards in time according to the SFHs from both the FAST fits in the 3D-HST catalogs and from \mname{}. Observed mass functions are taken from \citet{tomczak14} at three different redshifts and evolved to $z=3$. Overall, \mname{} provides much more realistic SFHs than FAST. The 3D-HST SFHs for low-mass galaxies are far too young at all redshifts while \prospector{} SFHs are slightly too old.
}
\label{fig:mf_evolve}
\end{sidewaysfigure*}

The results of this exercise are shown in Figure \ref{fig:mf_evolve}. For all combinations of $z_{\mathrm{start}}$ and $z_{\mathrm{obs}}$, the FAST SFHs greatly underpredict the number density of low-mass galaxies ($\lesssim 10^{10}$ M$_{\odot}$). This suggests that the exponentially declining SFHs assumed in FAST greatly underestimate the ages of low-mass galaxies, in agreement with the findings of \citet{wuyts11a} who use a similar methodology. Meanwhile, the predictions from the \mname{} SFHs are in much better agreement with the observations, though there are hints that there is more rapid evolution at higher redshifts ($z>2.5$) than predicted from the \mname{} SFHs.

The story is more complex at the higher masses. The 3D-HST SFHs underpredict the ages of massive galaxies at lower redshifts ($z\sim0.6$) but give much more accurate ages at $z\sim1.1,1.6$. The \mname{} SFHs accurately predict the evolution of very massive galaxies (M$_* > 10^{11}$ M$_{\odot}$), but somewhat overpredict the ages of galaxies around the knee of the mass function ($10^{10} <$ M$_*$/M$_{\odot} < 10^{11}$). 

In summary, the \mname{} SFHs present a remarkable improvement over the FAST SFHs, but there remain specific mass and redshift regimes which can be improved. The continuity prior appears to be a reasonable prior for some (even most) combinations of redshift and mass, but perhaps can be improved upon for galaxies around the knee of the mass function. A hierarchical Bayesian model would be a logical next step to craft an SFH prior which is simultaneously consistent with the observed SEDs and with observations such as the evolution of the stellar mass function with time. Thorough comparisons of SFHs derived from integrated light to those derived from resolved (e.g., \citealt{johnson13}) and semi-resolved \citep{cook19} stellar populations in local galaxies is also a promising way forward.

\subsection{A new consistency between independent inferences of the cosmic star formation rate density}
\label{sec:sfrd}
The cosmic star formation rate density is the rate of new stars produced per unit volume and unit time. In principle this quantity can be inferred with SED modeling in two ways: (1) by summing the instantaneous star formation rate for all galaxies in a fixed volume, or (2) by measuring the change in total stellar mass in the galaxy population as a function of time. Previous work has demonstrated that these two methods are inconsistent with one another at roughly the 0.3 dex level \citep{madau14,leja15,tomczak16}. While this offset is improved from the 0.6 dex discrepancy measured just a decade ago \citep{wilkins08}, it remains a serious concern as it implies systematic, across-the-board errors in inferred stellar masses and star formation rates at the factor-of-two level. Here we show that the new masses and star formation rates estimated with \mname{} resolve this tension.

We estimate $\rho_{\mathrm{SFR}}(z)$ (i.e., the SFRD) by again using the phenomenological description of the \citet{tomczak14} mass functions from \citet{leja15} as an intermediate step. This mass function is multiplied by SFR$_{\mathrm{Prospector}}$(M$_{\mathrm{FAST}}$) to produce the number density of galaxies as a function of SFR. The average value of SFR$_{\mathrm{Prospector}}$(M$_{\mathrm{FAST}}$) is calculated by stacking individual galaxy posterior PDFs for this quantity. This produces number density of galaxies as a function of SFR, which is then integrated numerically to produce the star formation rate density $\rho_{\mathrm{SFR}}$(z). This calculation is performed in small $\delta z$ steps between $0.5 < z < 2.5$. This procedure is repeated for \sfruvir{}. The integration is performed at a fixed mass range of $9 <$ log(M$_{\mathrm{FAST}}$/M$_{\odot}) < 13$ for all redshifts. 

To estimate $\dot{\rho}_{\mathrm{mass}}(z)$ (i.e., the SFRD from stellar mass growth), we take Equation 5 from \citet{tomczak14} describing the growth of stellar mass density from FAST:
\begin{equation}
\log(\rho_{\mathrm{mass}}) = a(1+z) +b
\end{equation}
with $\rho_{\mathrm{mass}}$ the total mass density in M$_{\odot}$/Mpc$^{3}$, $a=-0.33$, and $b=8.75$. The \mname{} stellar mass density is calculated using a correction to this equation estimated from M$_{\mathrm{Prospector}}$(M$_{\mathrm{FAST}}$) and the \citet{tomczak14} stellar mass functions. The stellar mass density $\rho_{\mathrm{mass}}$(z) is then converted into $\dot{\rho}_{\mathrm{mass}}$(z) by numerically estimating d$\rho_{\mathrm{mass}}$/dt between timesteps and multiplying by $1-R$, where $R$ is the fraction of mass ejected from a stellar population during the course of passive stellar evolution. This mass loss is assumed to occur instantaneously. For a \citet{chabrier03} IMF, $R=0.36$ \citep{leja15}.

This exercise produces $\dot{\rho}_{\mathrm{mass}}$ and $\rho_{\mathrm{SFR}}$ at $0.5 < z < 2.5$ from both \mname{} and from the combination of FAST stellar masses and \sfruvir{}. In principle, $\dot{\rho}_{\mathrm{mass}}$ and $\rho_{\mathrm{SFR}}$ may disagree when using a fixed mass selection as done in this work because both mass growth and star formation occurs in lower-mass galaxies.

\onecolumngrid
\clearpage
\onecolumngrid
\clearpage
\twocolumngrid

To correct for this effect we estimate the fraction of $\dot{\rho}_{\mathrm{mass}}$ and $\rho_{\mathrm{SFR}}$ occurring below this mass limit using the Universe Machine \citep{behroozi19}, a semi-empirical model which generates self-consistent estimates of the mass assembly history of the galaxy population. We caution that the estimated mass and SFR completeness estimated by comparing our sample selection criteria to the full 3D-HST catalog (Figure \ref{fig:coverage}) are slightly larger ($<10\%$) than those estimated from the Universe Machine. These completeness corrections have a limited effect on the comparison with previous SFR and mass density measurements, which is the key result of this paper, and so they are not explored further in this work. However, completeness corrections are essential to accurately measuring the total star formation rate density and will be derived self-consistently in future work.

\begin{figure*}
\begin{center}
\includegraphics[width=0.8\linewidth]{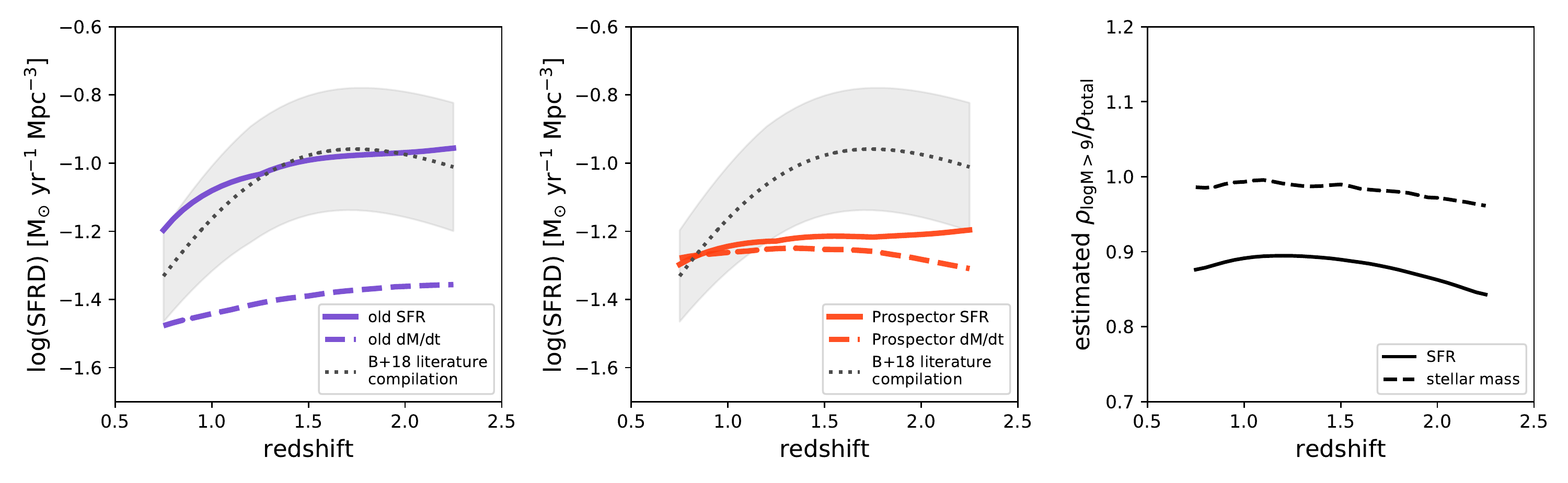}
\caption{Comparison between the observed cosmic SFRD and the cosmic SFRD implied by the observed growth of stellar masses. The canonical values from the FAST SED fitting code and \sfruvir{} (left panel) disagree such that there is too much observed star formation by $\sim0.2-0.4$ dex. The revised estimates from \mname{} (middle panel) largely remove this offset, due to a combination of lower star formation rates ($\sim0.2$ dex) and higher stellar masses ($\sim 0.1$ dex). As only galaxies with log(M$_{\mathrm{FAST}}$/M$_{\odot}) >9$ are modeled, a correction is applied for the stellar mass growth and star formation which occurs in galaxies with log(M$_*) < 9$. This correction factor is measured from the Universe Machine \citep{behroozi19} and shown in the right panel.}
\label{fig:madau}
\end{center}
\end{figure*}

The values of $\rho_{\mathrm{SFR}}$/$\dot{\rho}_{\mathrm{mass}}$ from these two procedures are shown in Figure \ref{fig:madau}. The combination of FAST dM/dt and \sfruvir{} recovers the inconsistency in SFRD inferences observed in previous work (e.g., \citealt{madau14,leja15,tomczak16}): a $\sim$0.3 dex gap between the observed SFRD and the SFRD implied by the mass function. Indeed, many galaxy formation models have long been in tension with the observed star formation rates at $1 < z < 3$, roughly at the factor of two level \citep{bouche10,firmani10,dave11,lilly13,dekel14,genel14,mitchell14}. Given that models of galaxy formation often calibrate themselves to the evolution of the stellar mass function, this tension is not unexpected \citep{leja15}.

This tension disappears with the new stellar masses and star formation rates from the \mname{} model. Internally, the star formation rate density decreases by $\sim$0.2 dex compared to \sfruvir{} while the observed growth of stellar mass increases by $\sim$0.1 dex compared to FAST stellar masses. The new estimates are internally consistent to within $\lesssim 0.1$ dex.

It is worth emphasizing that \prospector{} infers masses and SFRs using the same physical model. This is in contrast to the 3D-HST catalog masses and SFRs which are estimated from models with different and conflicting physical assumptions. It is better to use self-consistent estimates of mass and SFR when possible (e.g., \citealt{driver18}). Despite the internal consistency enforced in \mname{}, there is no guarantee that the {\it global} average of the stellar mass growth and star formation rate will agree. This makes the global $\lesssim0.1$ dex agreement quite remarkable.

\subsection{Comparison to dynamical masses}
\begin{figure*}[h!t!]
\begin{center}
\includegraphics[width=0.95\linewidth]{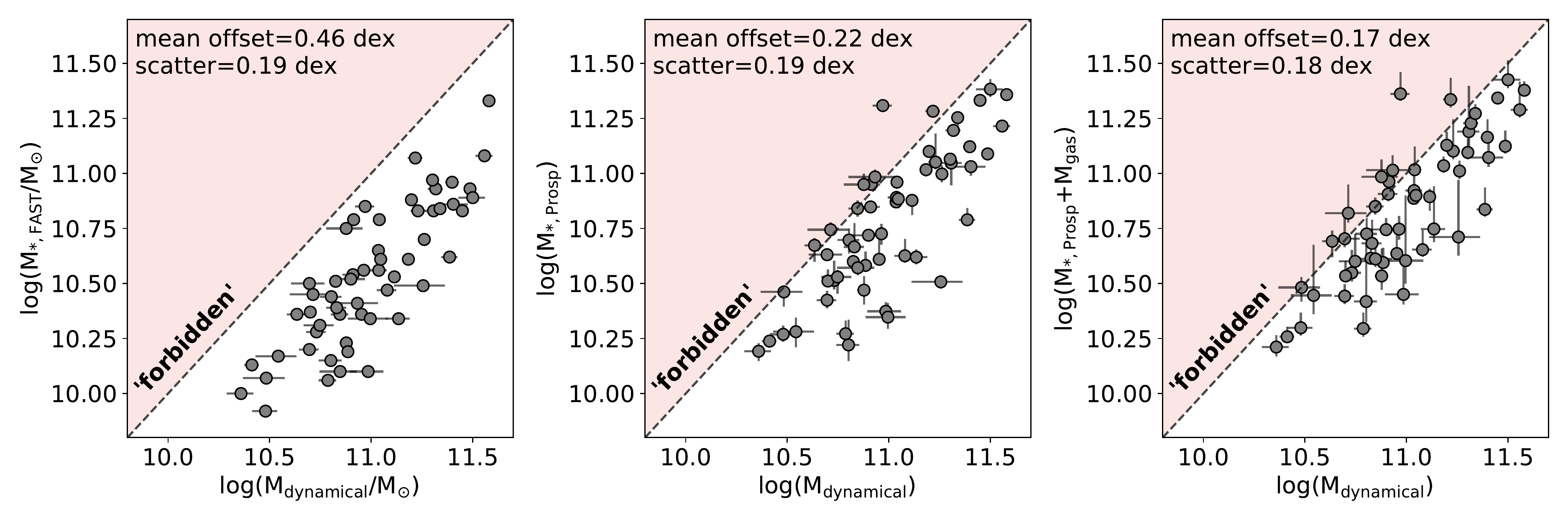}
\caption{Comparison between stellar and dynamical masses. The left panel shows stellar masses from FAST while the middle shows stellar masses from \mname{}. The scatter is similar though the offset decreases by $\sim$0.25 dex. The right panel includes molecular gas masses estimated from the scaling relationships of \citet{tacconi18}. The outlier in the `forbidden' region of the middle panel has a poorly determined stellar mass and is consistent with the dynamical constraint at the 3$\sigma$ level.}
\label{fig:dyncomp}
\end{center}
\end{figure*}

Galaxy dynamical masses are an independent constraint on stellar masses. More specifically, since the total galaxy mass budget is comprised of gaseous, stellar, and dark matter components, dynamical mass can be thought of as an ``upper limit'' to the stellar mass. Given that the \mname{} model increases stellar masses by an average of $\sim 0.2$ dex, it is important to ensure that the higher stellar masses do not violate dynamical constraints.

We test this with dynamical masses measured from deep Keck-DEIMOS spectra of star-forming and quiescent galaxies at $z\sim0.7$ \citep{bezanson15}. We adopt the structure-corrected dynamical masses calculated with the S\'ersic-dependent virial constant from \citet{cappellari06}. The dynamical masses are measured within the effective radius for each galaxy. We match 56 galaxies in the \citet{bezanson15} sample to the 3D-HST photometric catalogs and fit these galaxies with \mname{} using the spectroscopic redshifts from \citet{bezanson15}.

Figure \ref{fig:dyncomp} compares the measured dynamical masses to FAST stellar masses from  \citet{bezanson15} and to \mname{} stellar masses. The mean log(M$_{\mathrm{dyn}}$/M$_*$) is 0.46 dex for FAST stellar masses and 0.22 dex for \mname{} stellar masses. The final panel includes molecular gas masses estimated with the scaling relationships from \citet{tacconi18}; this has a small overall effect as most of the galaxies in this sample are quiescent. Crucially, this Figure demonstrates that the distribution of \mname{} masses do not violate the observed dynamical constraints. There is one object which is more massive than the dynamical constraints by $\sim$0.35 dex: however, it is consistent with the dynamical mass at the 3$\sigma$ level due to a long, non-Gaussian tail in the stellar mass posterior PDF.

While there is considerable scatter in M$_{\mathrm{Prospector}}$/M$_{\mathrm{FAST}}$, this scatter is not applied randomly as it seems to respect the dynamical constraints. This is unlikely to occur due to chance: using the observed distribution of M$_{\mathrm{Prospector}}$/M$_{\mathrm{FAST}}$ and applying these offsets randomly to M$_{\mathrm{FAST}}$ shows that 98\% of the time there should be additional critical outliers ($>$ 0.1 dex mass discrepancy) than the single one observed here. This implies that the increased stellar mass inferred by \mname{} is not added randomly, but instead is likely to reflect real variations in the underlying physical properties of these galaxies.

Overall, these results demonstrate that the new \mname{} stellar masses are consistent with the direct dynamical constraints. The new masses do leave less room on average for additional massive components such as dark matter or a more bottom-heavy IMF. A key question is whether the maximal allowed dark matter fractions are ``reasonable'' compared to hydrodynamical simulations of ellipticals and spirals. At these redshifts and masses, the Illustris TNG simulation suggests that dark matter should constitute about 50\% of the total matter within the effective radius \citep{lovell18}. This is closer to the revised stellar masses than the old stellar masses. Observational estimates of dark matter fractions necessarily rely on other methods to estimate stellar masses and in general create mixed expectations the amount of dark matter within the effective radius. For example, \citet{genzel17} finds that star-forming galaxies at $0.9<z<2.4$ have dark matter fractions of $<0.22$, but \citet{tiley19} argues that these should be considerably larger after correcting details of normalization prescription (they report dark matter fractions of $>60\%$ within 6 disk radii). \citet{cappellari13} use a variable IMF and measure dark matter fractions $<0.4$ in local early-types from the ATLAS-3D project. Ultimately, it is clear that the \mname{} masses are consistent with the dynamical masses in the sense that the stellar mass alone does not violate the constraints: however, given uncertainties in dynamical masses and expected galaxy-to-galaxy scatter in dark matter fractions, it remains to be seen whether the \mname{} masses are consistent with the full mass budget including dark matter.

\section{Discussion}
\label{sec:discussion}
The accuracy of the updated physical parameters presented in this work are necessarily contingent on the accuracy of the 14-parameter \mname{} model. Yet it can be challenging to perform hypothesis testing for high-redshift galaxy SED modeling due to the large number of ``unknowns'' relative to ``knowns''. We first discuss the necessity of performing model cross-validation to further verify, dismantle, or alter the new picture presented in this work (Section \ref{sec:falsifiability}). We then discuss potential future improvements in SED modeling which could further improve our interpretation of the observed galaxy photometry (Section \ref{sec:improvements}).
\subsection{Complex models and falsifiability}
\label{sec:falsifiability}
\begin{figure*}
\begin{center}
\includegraphics[width=0.98\linewidth]{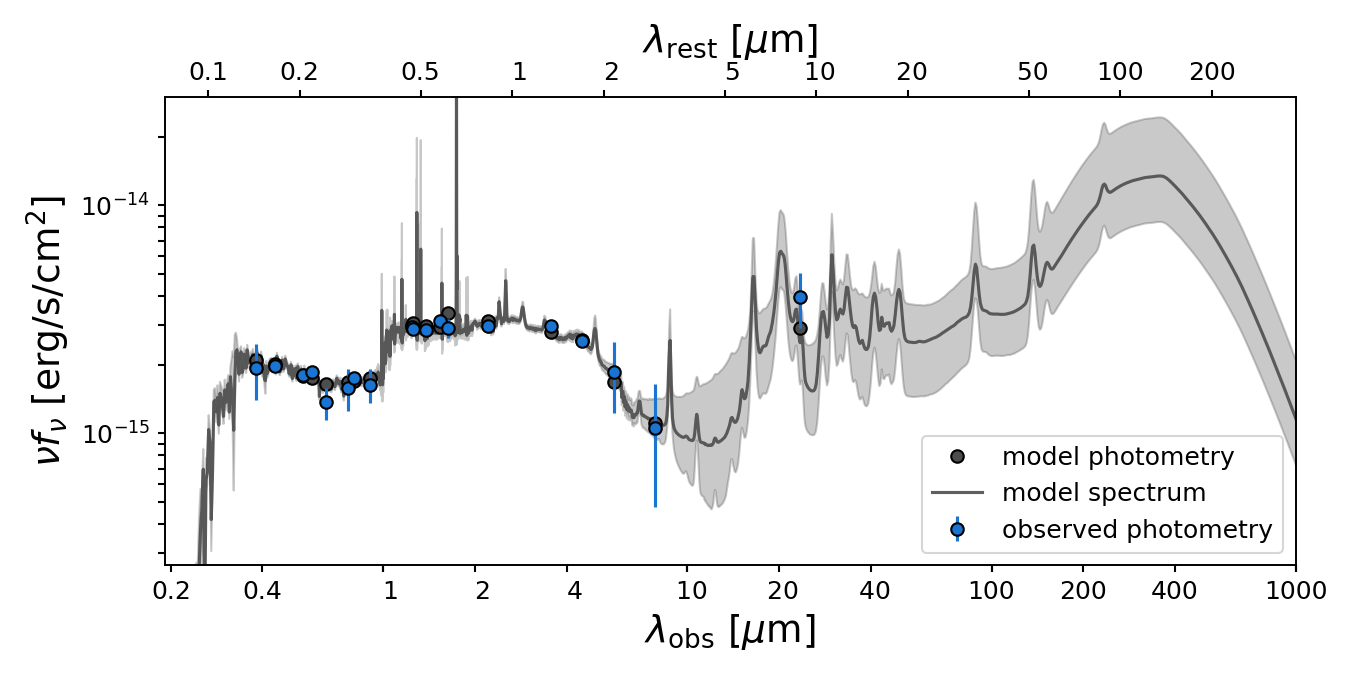}
\includegraphics[width=0.245\linewidth]{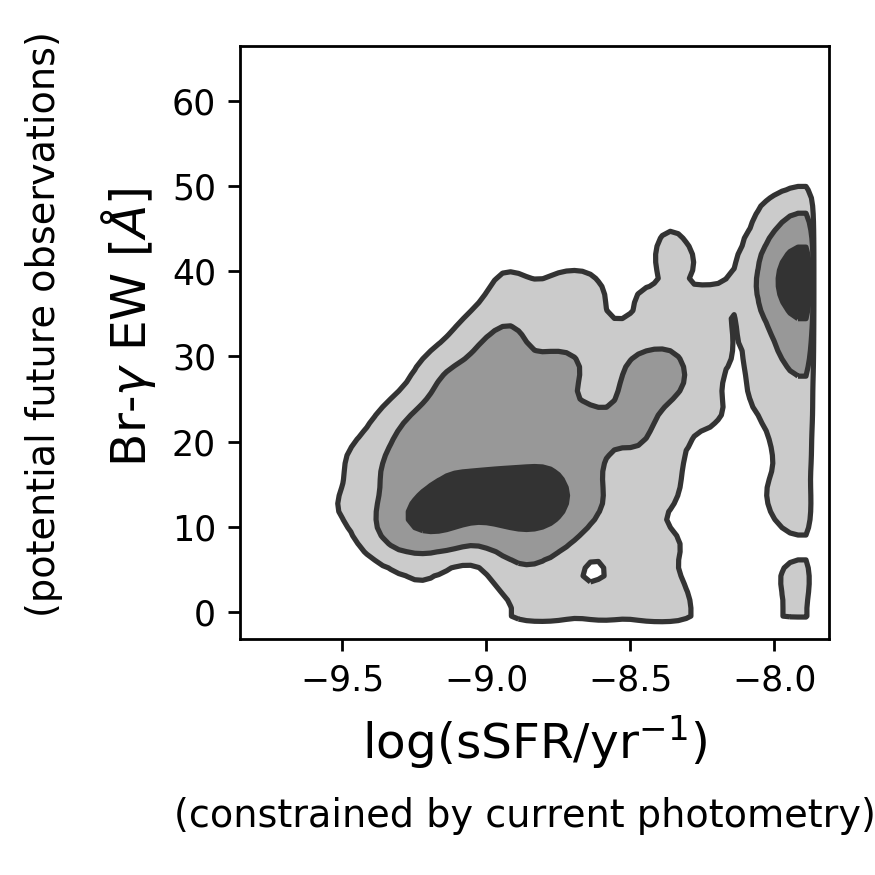}
\includegraphics[width=0.245\linewidth]{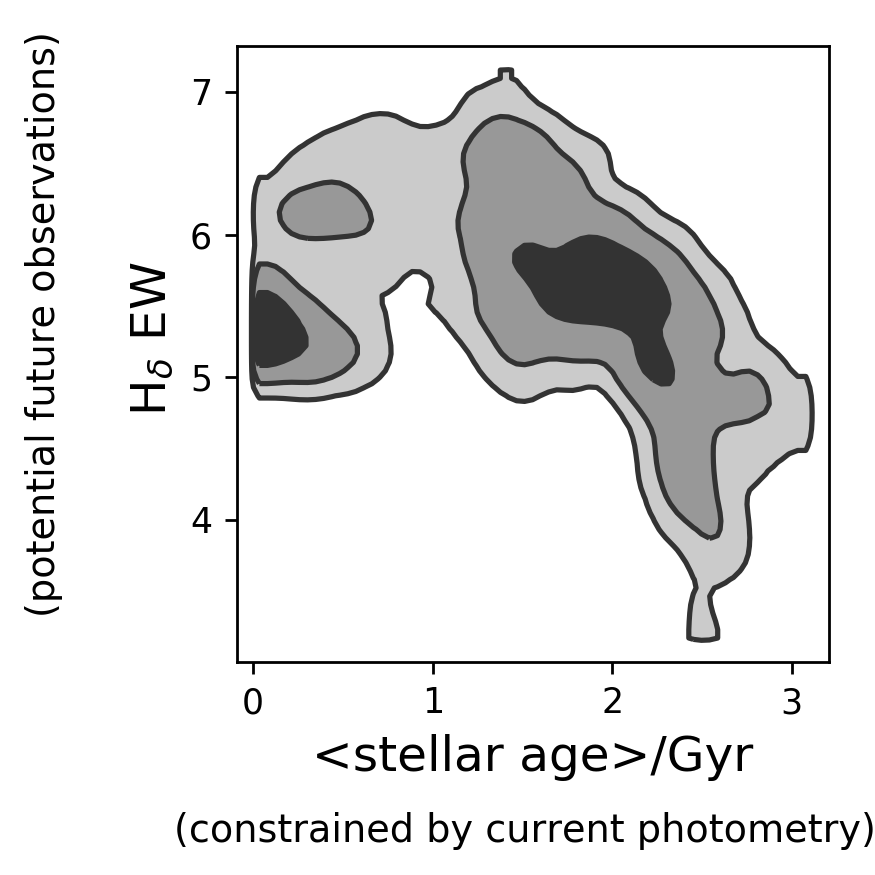}
\includegraphics[width=0.245\linewidth]{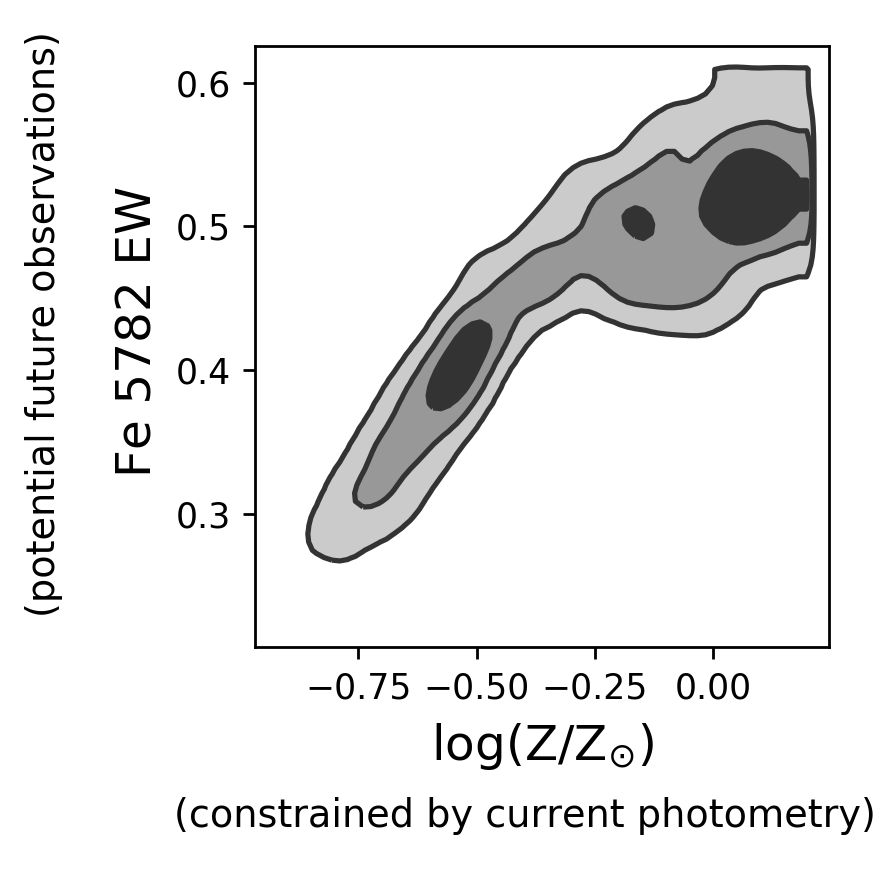}
\includegraphics[width=0.245\linewidth]{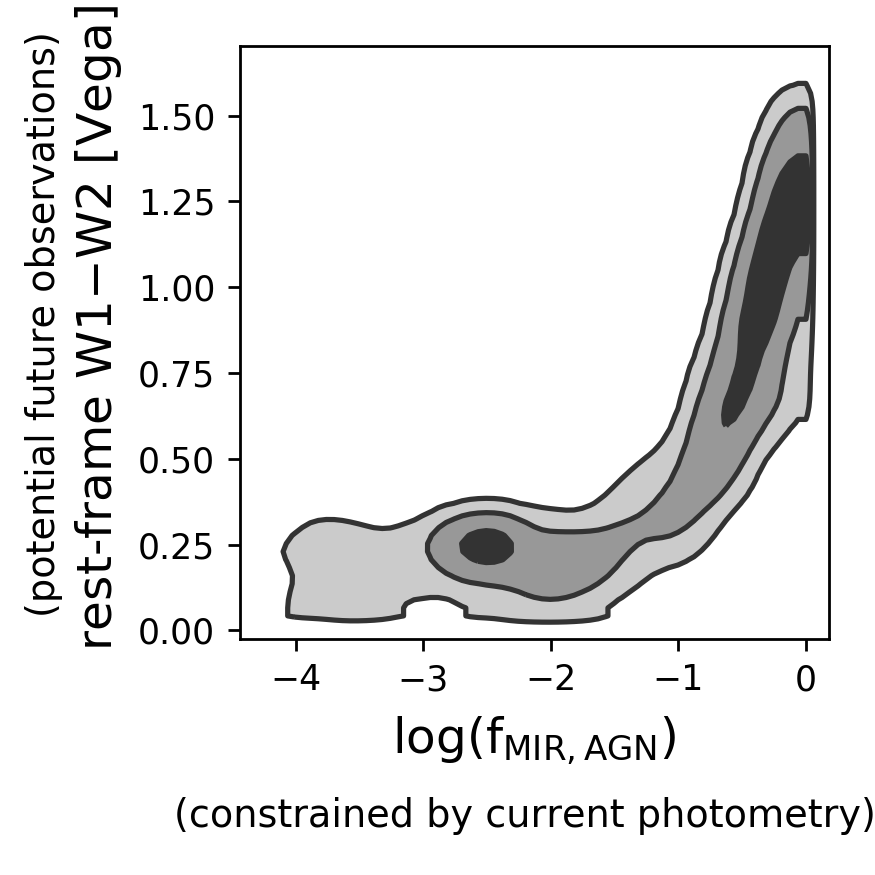}
\caption{Future data has the potential to better constrain parameters in the \mname{} model. The top panel shows the fit to the photometry of a galaxy from the 3D-HST catalogs, UDS 7610. The grey shaded region in the upper panel represents the 1-sigma range of model spectra drawn from the posteriors. The lower panels show predictions for future data which can constrain the major uncertainties in the \mname{} posteriors. The shaded regions in the lower panels correspond to 1, 2, and 3$\sigma$ ranges.}
\label{fig:observables}
\end{center}
\end{figure*}
In this work we present new inferences of stellar masses and SFRs from a high-dimensional physical model for galaxy SEDs. This model pushes the field forward by allowing galaxy-to-galaxy variation for many components of galaxy formation which were fixed in previous work, such as the shape of the dust attenuation curve or the highly flexible step-function SFHs. This is possible because of advances in statistical and sampling methodologies, the ongoing and dramatic decrease in the price of computing time, and substantial improvements in stellar population synthesis techniques.

The primary challenge in evaluating this model (or any such model) is that there is no ``ground truth" with which to compare basic properties derived from galaxy SED fitting. Due to this lack of corroboration, there has been a long history of skepticism in the literature about the accuracy of galaxy SED modeling results (e.g., \citealt{papovich01,shapley05,conroy09,wuyts09,behroozi10,walcher11,taylor11,mobasher15,santini15}). 

Fitting simulated galaxies with galaxy SED models is a useful way to cross-examine their assumptions (e.g., \citealt{hayward15}), as this is a scenario in which the ground truth is known. Simulations reproduce many key components of galaxy formation, including complex star formation histories, physically motivated metallicity enrichment histories, and (for high-resolution simulations) complex spatial mixtures of stars and dust. However, such comparisons are only useful insofar as the physical conditions in simulated galaxies approximate those of real galaxies. It has been shown that the outputs of numerical simulations of galaxy formation are sensitive to the implementation of their sub-grid physics (e.g., \citealt{crain15}). This is notable because different numerical simulations adopt different sub-grid physics recipes \citep{somerville15}. This means the accuracy of simulation outputs vary according to the accuracy of their unique sub-grid recipes, which are difficult to assess. Furthermore, it is only possible to use simulations to test SED fitting ingredients which are {\it not} inputs to simulated galaxies. This forbids testing many basic components of galaxy SED models, including stellar population synthesis assumptions, AGN emission models, and the sub-resolution behavior of dust and the interstellar medium \citep{smith15,nelson18}.

Given that a direct comparison between SED modeling results and ground truth is not possible, we suggest here that next best approach is to build a model which is, to the greatest extent possible, consistent with all other observations. This involves projecting the implications of galaxy SED models conditional on the observed data into the space of completely independent observables. Informative comparisons of this type can include comparing stellar masses to dynamical masses \citep{erb06b,taylor12}, predicting the strength of spectral features from fits to the photometry \citep{leja17}, and comparing star formation histories of galaxies at low redshift to the observed star formation rates and stellar masses of galaxies at higher redshift \citep{wuyts11b}. This approach is particularly fruitful for galaxy SED fitting: due to the covariance of basic parameters like age, dust, and metallicity, a simple change to the prior for one parameter can have ramifications for many other parameters of interest.

Figure \ref{fig:observables} illustrates the potential for additional data to further constrain the parameters in the \mname{} model. The top panel shows a model fit to photometry from the 3D-HST survey. The lower panels show the joint PDF between key model parameters (specific star formation rate, AGN strength, stellar metallicity, and stellar age) and potential future observables (Br-$\gamma$ emission equivalent width, H$\delta$ and Fe 5782$\AA$ absorption equivalent width, and {\it WISE} rest-frame mid-infrared colors). The covariance between these parameters means that future observations can constrain key remaining uncertainties in the \mname{} models. Notably, while these types of covariances are very common, the particular galaxy shown in Figure \ref{fig:observables} is unusual in displaying strong covariances with all of these observables at once.

\subsection{Towards a more accurate SED model}
\label{sec:improvements}
One key improvement in \mname{} is the large number of free parameters coupled with the statistical machinery to put realistic constraints on them. Allowing significant deviations from the ``standard script'' for galaxy formation permits more accurate properties to be inferred on a galaxy-by-galaxy basis.

However, there are still a number of key physical parameters which remain fixed. It is reasonable to think of models such as \mname{} as one important step towards the ultimate goal, which is a fully flexible physical model for galaxy emission across all redshifts. One key step is to constrain the stellar attenuation as a function of age in order to confirm the global effect of old stellar heating as described in Section \ref{sec:oldem}. Here we discuss several additional future steps on the path to this goal.

\subsubsection{Propagation of redshift uncertainties}
\label{sec:zerr}
\mname{} treats redshift as a fixed parameter. This approach explicitly neglects the effect of errors in distance determination on the resulting galaxy properties. 

This assumption will affect some galaxy fits more than others. In the 3D-HST catalogs, redshift has been inferred independently from a combination of $HST$ grism spectroscopy, ground-based spectroscopy, and photometric redshifts from EAZY.  A fixed redshift is an excellent approximation for galaxies with solid spectroscopic or grism redshifts but is a less robust approximation for photometric redshifts. The reliability of photometric redshifts will also scale with the signal-to-noise of the photometry. For example, the scatter between photometric and spectroscopic redshifts for the entire 3D-HST survey is 0.0197, but for galaxies with H$_{F160W}$ magnitude  $>26$ this scatter increases to $\sim$0.05 \citep{bezanson16}.

Redshift errors can have a strong effect on the physical parameters inferred from SED fitting. For example, \citet{chevallard16} use the Bayesian SED fitting tool BEAGLE to fit two high-redshift galaxy SEDs simultaneously for redshift and stellar populations parameters. The results show that redshift can have a complex interplay with the derived stellar populations parameters: even moderate redshift errors of $\sim$0.15 can affect individual stellar masses by a full order of magnitude or more. The systematic effect of redshift errors on global properties of the galaxy population -- such as the stellar mass function or the cosmic star formation rate density -- has yet to be characterized in a Bayesian framework.

One simple step way forward is to use posteriors from photometric and grism redshift-fitting codes as priors for the redshift estimated in SED fitting. This is an imperfect solution, as it mixes multiple different assumptions about stellar populations. Ultimately, it would be ideal to use a single workflow to analyze all the available spectroscopic and photometric data and then simultaneously estimate redshifts and stellar populations parameters: for a first step towards this, see \citealt{acquaviva15}.

\subsubsection{A flexible IR SED}
\label{sec:fixedirsed} 
In this work we adopt a fixed shape for the IR SED. This fixed shape is used to extrapolate the total infrared luminosity from the observed MIPS 24$\mu$m photometry. The total infrared luminosity is a critical parameter as it is closely related to the total star formation rate, though the exact relationship depends on the stellar properties as discussed in Section \ref{sec:oldem}. Approximating the IR SED as fixed is helpful due to the lack of MIR or FIR photometry at intermediate and high redshifts for the majority of the galaxy population. However, the IR SED shows significant variation on a galaxy-by-galaxy basis in the local Universe (e.g., \citealt{dale05,dale12}) and this variation is likely to persist at higher redshifts.

Observations of variations in IR SED shape at higher redshifts are limited by the depth of available $Herschel$ photometry. The handful of galaxies with individual detections in $Herschel$ IR photometry show that the $L_{8\mu \mathrm{m}}$/\lir{} ratio has a scatter of a factor of $\sim$2, with a tail towards higher values of $L_{8\mu \mathrm{m}}$/\lir{} in systems with SFR$\gtrsim$ 100 M$_{\odot}$/yr \citep{elbaz11,wuyts11a}. Lower flux limits can be reached with stacking analysis. \citet{shivaei17} show that the $L_{8\mu \mathrm{m}}$/\lir{} conversion is likely a strong function of stellar mass as well, varying $systematically$ by a factor of 2 when comparing massive galaxies to galaxies with log(M/M$_{\odot}$) $<$ 10. $L_{8\mu \mathrm{m}}$/\lir{} also shows significant redshift evolution \citep{whitaker17}.

A comprehensive study of the variation of the IR SED at $z>0.5$ with galaxy properties has not yet been performed due to the shallow limits of the available MIR and FIR imaging. Stacking or deblending $Herschel$ photometry combined with accurate galaxy properties from SED modeling is one potential way to address this issue. It would be straightforward to incorporate these results into galaxy SED fitting models via priors. Systematic change in the IR SED with galaxy properties has the potential to alter important galaxy scaling relationships such as the low-mass slope of the star-forming sequence in a mass-dependent fashion and correspondingly alter the cosmic star formation rate density \citep{whitaker14,leja15,shivaei17}.

\subsubsection{$\alpha$-element abundances}
Galaxy SED models currently assume a solar abundance pattern by necessity. However, there is clear evidence from high-resolution spectra of quiescent galaxies that the $\alpha$-element abundance varies systematically with galaxy properties. This correlation is apparent in the nearby Universe where massive galaxies have [$\alpha$/Fe] $\sim$ +0.23 \citep{thomas05,conroy14}. This trend increases strengthens at intermediate redshifts ($0.5 < z < 2$), where small samples of massive galaxies have [$\alpha$/Fe] $\sim$ +0.3 \citep{choi14,onodera15}. More extreme individual causes have been detected, including [$\alpha$/Fe] $>$ 0.4 \citep{lonoce15} and [$\alpha$/Fe] = +0.6 \citep{kriek16}. It is more difficult to infer elemental abundance patterns in young galaxies due to the lack of strong absorption lines, but simulations predict that star-forming galaxies have trends in $\alpha$-element abundance patterns with mass, redshift, and star formation rate \citep{matthee18}. These can be [$\alpha$/Fe] = +0.6 or higher in highly star-forming galaxies at $z=2$ and above, consistent with observed nebular abundances \citep{steidel16}.

These trends in abundance patterns have ramifications for the integrated photometry of galaxies. \citet{vazdekis15} generate $\alpha$-enhanced models with [$\alpha$/Fe] = +0.4 and show that the resulting optical fluxes change by 10\%-40\% and the optical colors change by $\sim$0.1 magnitude, depending on the age and metallicity of the stars. This suggests that variations in $\alpha$-element patterns should be included when fitting galaxy photometry: for example, $\alpha$ abundance patterns could be important in explaining the $ugr$ colors of massive ellipticals, which have been too red in models for many years (e.g., \citealt{conroy10,vazdekis15}). \citet{choi19} show explicitly that synthesizing $ugriz$ fluxes from the best-fit spectrum with individual elemental abundances allowed to vary will reproduce the observed colors to within $<0.03$ magnitudes, while using solar-scaled abundances results in larger residuals (up to 0.1 magnitudes for the oldest systems).

It remains unclear how much variation in $\alpha$-element abundance will affect SED modeling at higher redshifts. On one hand the $\alpha$-element abundance patterns are more extreme at higher redshifts, but on the other hand galaxies are younger on average and therefore less sensitive to $\alpha$-element variations. Future versions of \fsps{} will include variation in the $\alpha$-abundance pattern, providing a straightforward way to include the effect of variations in $\alpha$-enhancement on galaxy properties derived from SED modeling.

\subsubsection{IMF Variations}
The shape of the stellar initial mass function is a critical assumption in galaxy SED modeling. Changing the IMF below $\sim0.8$M$_{\odot}$ substantially changes inferred stellar masses and star formation rates without affecting significantly changing the predicted SED. Such a change does not affect the global agreement between the SFRD and the growth of the stellar mass density as both SFR and mass are changed proportionally \citep{leja15}. Changing the high-mass end of the IMF will substantially change the inferred SFRs again without much consequence for the predicted SEDs, though this change would alter the global agreement between mass and SFR.

Recent work has provided solid evidence in nearby galaxies for long-suspected systematic variations in the stellar IMF between galaxies. IR spectroscopy \citep{vd10b,conroy12b}, dynamical modeling \citep{cappellari13}, and gravitational lens analysis \citep{treu10} all independently suggest that ellipticals with higher velocity dispersions have increasingly `bottom-heavy' IMFs, though there remains some tension in the exact agreement between these techniques \citep{newman17}. Star counts in ultra-faint dwarf galaxies find that these galaxies are deficient in low-mass stars (`bottom-light') \citep{geha13}. These results taken together are qualitatively consistent with a continuous variation in the IMF from low-mass to high-mass galaxies. 

The ramifications of a variable IMF for the $z\gtrsim1$ galaxy population have not been fully explored. This is at least in part because of the paucity of observables which directly correlate with the IMF \citep{vd12}. There is also recent evidence that bottom-heavy IMFs might only be confined to the very central regions \citep{vd17,conroy17,sarzi18,zhou19}, making comparisons to global quantities challenging. This results in a greater emphasis on indirect methods such as comparing the inferred stellar and dynamical masses. These comparisons typically assume canonical IMFs and find that the stellar mass takes up an increasing fraction of the total mass budget at higher redshift \citep{vandesande13,belli17}, or even exceed the total mass budget \citep{price19}. This presents a difficult conundrum: if old galaxies in the local Universe show bottom-heavy IMFs, why are these IMFs seemingly incompatible with dynamical measurements of their putative progenitor galaxies at $z\sim2$? One intriguing possibility is that the star-forming progenitors of the cores of local elliptical galaxies have yet to be found due to high levels of dust attenuation \citep{nelson14}.

Making progress on this issue will require careful simultaneous dynamical and SED modeling in order to satisfy both dynamical constraints and the observed photometry.  Such work will be crucial to ensuring the absolute accuracy of SED-derived quantities.

\section{Conclusions}
\label{sec:conclusion}
In this work we present a revised estimate on the rate of galaxy stellar mass assembly at $0.5 < z < 2.5$ using the \mname{} galaxy physical model. The primary advance over previous work is the much larger number of physical parameters which are modeled within \prospector{} ($N=14$, compared to $N\sim 4-7$). This high dimensionality permits modeling the effect of a number of second-order physical effects on both stellar mass and SFR estimates on an object-by-object basis. These new high-dimensional SED models are possible due to a number of technical improvements: the nested sampling routine \dynesty{}, on-the-fly model generation with \fsps{}, and the \prospector{} Bayesian inference framework.

We fit a version of the \mname{} physical model from \citet{leja17,leja18} modified for high-redshift galaxies. This model makes use of the wide range of physics available in \fsps{} and has a total of 14 free parameters. These physics include a flexible 6-parameter nonparametric SFH, state-of-the-art MIST stellar isochrones, a broad range of stellar metallicities, a two-component dust attenuation model with a flexible dust attenuation curve, dust emission via energy balance, nebular line and continuum emission, and a model for the MIR emission of dusty AGN torii.

The \mname{} model is fit to rest-frame UV-MIR photometry of 58,461 galaxies from the 3D-HST survey in the redshift range $0.5 < z < 2.5$. These catalogs provide between an immense amount of information: there are between 17 and 44 bands of aperture-matched photometry available across 5 distinct extragalactic fields. These photometric data are coupled with redshifts inferred from a combination of ground-based spectroscopy, the $HST$ G141 grism, and photometric redshifts from EAZY. After fitting these data, we present the following conclusions:

\begin{enumerate}
\item[1a.] The \mname{} stellar masses are systematically $0.1-0.3$ dex higher than stellar masses from the 3D-HST catalogs inferred with the FAST SED fitting code. This offset correlates with stellar mass and, more weakly, with redshift. 
\item[1b.] While multiple effects contribute at a low level, the primary cause of the offset is the older stellar ages inferred with \mname{}. Comparing stacked SFHs inferred from the 3D-HST SED and the \mname{} model show that these differences can be dramatic: highly star-forming galaxies are older by a factor of $\sim 10$ and galaxies on the star-forming sequence are older by a factor of $\sim 5$.
\item[2a.] The \mname{} star formation rates match state-of-the-art UV+IR SFRs at high sSFRs (log(sSFR/yr$^{-1})\approx8$). They are increasingly lower than \sfruvir{} with decreasing sSFR such that by log(sSFR/yr$^{-1})\approx-10.5$ there is an offset of $0.75-1$ dex. 
\item[2b.] While again multiple effects contribute, the largest cause of this offset is the emission from from old stars. This is neglected in \sfruvir{} but self-consistently estimated in the \mname{} model. The fraction of \lir{}+\luv{} powered by emission from `old' ($t>100$ Myr) stars as a function of sSFR is derived and an equation to estimate this effect is presented.
\end{enumerate}

We explore the global implications of these new inferences with several model cross-validation techniques:
\begin{enumerate}
\item[i.] The global star formation rate density is estimated from the SED fits using both dM$_*$/dt and SFR(t). These two estimators are inconsistent when estimated with FAST stellar masses and \sfruvir{} in the sense that $\rho_{\mathrm{SFR}}$ is higher than $\dot{\rho}_{\mathrm{mass}}$ by $\sim 0.3$ dex, in agreement with other studies in the literature. The \mname{} estimates bring $\rho_{\mathrm{SFR}}$ down by $\sim$0.2 dex and $\dot{\rho}_{\mathrm{mass}}$ up by $\sim$0.1 dex such that there is now consistency in the inferred SFRD. This is a notable finding as there is no guarantee of self-consistency in the cosmic sum of these values.
\item[ii.] The \mname{} SFHs are much better predictors of the redshift evolution of the stellar mass function. This is demonstrated by using observed star formation histories coupled with a merger model to wind the observed stellar mass function back in time. This model mass function is compared to the observed stellar mass functions to test the consistency of the SFHs. The \mname{} SFHs are older on average and better describe the observations than the 3D-HST SFHs across most combinations of mass and redshift, though galaxies in the knee of the mass function ($10 < $ log(M/M$_{\odot}) < 11$) are likely too old within the \mname{} model.
\item[iii.] The new stellar masses from \mname{} are consistent with observed dynamical constraints, with the average offset between stellar and dynamical mass decreasing from $\sim0.46$ dex to $\sim 0.22$ dex (0.17 dex when including gas), though the new masses do leave less room on average for additional components such as dark matter or a more bottom-heavy IMF.
\end{enumerate}

The primary goal of this work is to build a model for galaxy properties which is, to the greatest extent possible, consistent with all observations. We take the first steps in this direction by performing cross-validation both within \mname{} and with external data sets and by highlighting future observations which will provide deeper constraints for the \mname{} physical model. Such future data will lead to updates of model priors used in SED fitting. Due to the covariance of basic galaxy parameters, a change to the prior for one parameter will have ramifications for other parameters of interest: in this way such updates will create ``evolving results'' . It is hoped that this methodology can be used to converge towards the truth. 

\acknowledgements J.L. is supported by an NSF Astronomy and Astrophysics Postdoctoral Fellowship under award AST-1701487. We thank Sandra Faber, Sandro Tacchella, Maarten Baes, and Rohan Naidu for fruitful discussions. The computations in this paper were run on the Odyssey cluster supported by the FAS Division of Science, Research Computing Group at Harvard University. This research made use of Astropy,\footnote{http://www.astropy.org} a community-developed core Python package for Astronomy \citep{astropy13, astropy18}.
\clearpage

\software{\texttt{Prospector} \citep{prospector17}, \texttt{python-fsps} \citep{pythonfsps14}, \texttt{Astropy} \citep{astropy13,astropy18}, \texttt{FSPS} \citep{conroy09b}, \texttt{matplotlib} \citep{matplotlib18}, \texttt{scipy} \citep{scipy}, \texttt{ipython} \citep{ipython}, \texttt{numpy} \citep{numpy}}

\bibliography{/Users/joel/my_papers/tex_files/jrlbib}
\end{document}